\documentclass[11pt,journal,twoside,onecolumn]{IEEEtran}

\usepackage{psfrag,epsfig,graphics}
\usepackage{amsmath,amssymb,multirow}
\usepackage[ruled,vlined]{algorithm2e}
\usepackage{subfigure}%,ulem}
  \usepackage[below]{placeins}
  \usepackage{cite}
    \usepackage{afterpage}

\usepackage{color}  % To include comments in color
\def\cred{\textcolor{black}}
\def\cblue{\textcolor{black}}
\def\cmag{\textcolor{black}}

\newcommand{\cb}[1]{{\boldsymbol{#1}}}
\newcommand{\cp}[1]{\ifmmode {\mathcal{#1}}\else ${\mathcal{#1}}$\fi}
\newcommand{\balpha}{\boldsymbol{\alpha}}

\newcommand{\bpsi}{\boldsymbol{\psi}}
\newcommand{\bzeta}{\boldsymbol{\zeta}}

\newcommand{\f}{\boldsymbol{f}}
\newcommand{\bu}{\boldsymbol{u}}
\newcommand{\bp}{\boldsymbol{p}}

\newcommand{\bg}{\boldsymbol{g}}
\newcommand{\br}{\boldsymbol{r}}

\newcommand{\bw}{\boldsymbol{w}}
\newcommand{\bx}{\boldsymbol{x}}
\newcommand{\by}{\boldsymbol{y}}
\newcommand{\bv}{\boldsymbol{v}}
\newcommand{\bW}{\boldsymbol{W}}

\newcommand{\bB}{\boldsymbol{B}}
\newcommand{\bC}{\boldsymbol{C}}
\newcommand{\bK}{\boldsymbol{K}}
\newcommand{\bQ}{\boldsymbol{Q}}
\newcommand{\bG}{\boldsymbol{G}}
\newcommand{\bP}{\boldsymbol{P}}
\newcommand{\bH}{\boldsymbol{H}}
\newcommand{\bA}{\boldsymbol{A}}
\newcommand{\bR}{\boldsymbol{R}}
\newcommand{\bV}{\boldsymbol{V}}
\newcommand{\bM}{\boldsymbol{M}}
\newcommand{\bY}{\boldsymbol{Y}}
\newcommand{\bSig}{\boldsymbol{\Sigma}}
\newcommand{\bsig}{\boldsymbol{\sigma}}

\newcommand{\bGam}{\boldsymbol{\Gamma}}
\newcommand{\bI}{\boldsymbol{I}}

\newcommand{\N}[1]{\cp{N}_{#1}}
\newcommand{\C}{\cp{C}}

\newcommand{\tr}{\text{trace}}
\newcommand{\vc}{\text{vec}}

\newtheorem{theorem}{Theorem}

\newtheorem{assumption}{Assumption}

\begin{document}

\title{Multitask Diffusion Adaptation over Networks}
\author{Jie Chen$^\dag$, \IEEEmembership{Student Member, IEEE}, C{\'e}dric Richard$^\dag$, \IEEEmembership{Senior Member, IEEE} \\
Ali H. Sayed$^\ddag$, \IEEEmembership{Fellow Member, IEEE}
\thanks{This work was partly supported by the Agence Nationale pour la Recherche, France, (Hypanema project, ANR-12-BS03-003), and the Centre National de la Recherche Scientifique, France (Display project, Mastodons). \cblue{The} work of A. H. Sayed was supported in part by NSF grant CCF-1011918.}
 \\ \vspace{0.5cm}
\small{\linespread{0.2} $^\dag$ Universit{\'e} de Nice Sophia-Antipolis, UMR CNRS 7293, Observatoire de la C{\^{o}}te d'Azur \\
Laboratoire Lagrange, Parc Valrose, 06102 Nice - France \\
phone: (33) 492 076 394 \hspace{0.5cm} \hspace{0.5cm} fax:
(33) 492 076 321 \\ jie.chen@unice.fr \hspace{0.5cm}
cedric.richard@unice.fr}
\vspace{0.3cm}\\
\small{\linespread{0.2} $^\ddag$ Electrical Engineering Department \\
University of California, Los Angeles, USA \\
phone: (310) 267 2142 \hspace{0.5cm} \hspace{0.5cm} fax:
(310) 206 8495 \\
sayed@ee.ucla.edu}
}

\maketitle

\vspace{-0.5cm}

\hspace{0.4cm}{\bf{\small EDICS:}}   \small{NET-ADEG, NET-DISP, MLR-DIST, SSP-PERF}

\vspace{1cm}

\begin{abstract}
Adaptive networks are suitable for decentralized inference tasks, e.g., to monitor complex natural phenomena. Recent research works have intensively studied distributed optimization problems in the case where the nodes have to estimate a single optimum parameter vector collaboratively. However, there are many important applications that are multitask-oriented in the sense that there are multiple optimum parameter vectors to be inferred simultaneously, in a collaborative manner, over the area covered by the network. In this paper, we employ diffusion strategies to develop distributed algorithms that address multitask problems by minimizing an appropriate mean-square error criterion with $\ell_2$-regularization. The stability and convergence of the algorithm in the mean and in the mean-square sense is analyzed.  Simulations are conducted to verify the theoretical findings, and to illustrate how the distributed strategy can be used in several useful applications related to spectral sensing, target localization, and hyperspectral data unmixing.
\end{abstract}


\newpage

\section{Introduction}

Distributed adaptation over networks has emerged as an attractive and challenging research area with the advent of multi-agent (wireless or wireline) networks. Accessible overviews of recent results in the field can be found in~\cite{Sayed2013diff,Sayed2013intr}. In adaptive networks, the interconnected nodes have to \cblue{continually learn and adapt, as well as} perform preassigned tasks such as parameter estimation from observations collected by the \cblue{dispersed agents}. Although centralized strategies with a fusion center \cblue{can benefit more fully}  from information collected throughout the network but stored at a single point, in most cases, distributed strategies are more attractive to solve \cblue{inference} problems in a collaborative and autonomous manner. Scalability, robustness, and low-power consumption are key characteristics of these strategies. Applications include environment monitoring, but also modeling of self-organized behavior observed in nature such as bird flight in formation and fish schooling~\cite{Sayed2013diff,Tu2011Mobile}.

There are several useful distributed strategies for sequential data processing over networks including consensus strategies~\cite{Tsitsiklis1984,Xiao2004,Braca2008,Nedic2009,Kar2009,Srivastava2011}, incremental strategies~\cite{Bertsekas1997,Nedic2001,Rabbat2005,Blatt2007,Lopes2007incr}, and diffusion strategies~\cite{Sayed2013diff,Sayed2013intr,Lopes2008diff,Cattivelli2010diff,ChenUCLA2012,ChenUCLA2013}. \cmag{Incremental techniques require the determination of a cyclic path that runs across the nodes, which is generally a challenging (NP-hard) task to perform. Besides, incremental solutions can be problematic for adaptation over networks because they are sensitive to link failures. On the other hand, diffusion strategies are attractive since they are scalable, robust, and enable continuous adaptation and learning.  In addition, for data processing over adaptive networks, diffusion strategies have been shown to have superior stability and performance ranges~\cite{Tu2012} than consensus-based implementations.} Consequently, we shall focus on diffusion-type implementations in the sequel. The diffusion LMS \cblue{strategy} was  proposed and studied in~\cite{Lopes2008diff,Cattivelli2010diff}. Its performance in the presence of imperfect information exchange and model non-stationarity was analyzed in~\cite{Tu2011,Khalili2012,Zhao2012impe}. Diffusion LMS with $\ell_1$-norm regularization was considered in~\cite{Lorenzo2012,Liu2012,Chouvardas2012,Lorenzo2013spar} to promote sparsity in the model. In~\cite{Chouvardas2011set}, the problem of distributed learning in diffusion networks was addressed by deriving projection algorithms onto convex sets. Diffusion RLS over adaptive networks was studied in~\cite{Cattivelli2008,Bertrand2011RLS}. More recently, a distributed dictionary learning algorithm based on a diffusion strategy was derived in \cite{chainais2013distributed, chainais2013learning,Wee2013}. This literature mainly considers quadratic cost functions and linear models where systems are characterized by a parameter vector in the Euclidean space. Extensions to more general cost functions that are not necessarily quadratic and to more general data models are studied in~\cite{ChenUCLA2012,ChenUCLA2013} in the context of adaptation and learning over networks. Moreover, several other works explored distributed estimation for nonlinear input-output relationships defined in a functional space, such as reproducing kernel Hilbert spaces. For instance, in~\cite{Predd2006WSN}, inference is performed with a regularized kernel least-squares estimator, where the distributed information-sharing strategy consisted of successive orthogonal projections. Distributed estimation based on adaptive kernel regression~\cite{richard2009online,honeine2007line} is also studied in~\cite{Honeine2010WSN,honeine2009functional,honeine2008regression,honeine2008distributed}, with information passed from node to node in an incremental manner. In~\cite{Chen2010WSN}, non-negative distributed regression is considered for nonlinear model inference subject to non-negativity constraints, where the  diffusion strategy is used to conduct information exchange. %Reproducing kernel Hilbert spaces were also considered for auto-localization of sensors in~\cite{essoloh2008distributed,honeine2008localization,essoloh2007anchor} by solving a pre-image problem~\cite{honeine2011preimage,honeine2009solving}.  

An inspection of the existing literature on distributed algorithms shows that most works focus primarily, though not exclusively~\cite{Tu2012Asilomar,Zhao2012,Bogdanovic2013}, on the case where the nodes have to estimate a single optimum parameter vector collaboratively. We shall refer to problems of this type as \emph{single-task}  problems. However, many problems of interest happen to be \emph{multitask}-oriented in the sense that there are multiple optimum parameter vectors to be inferred simultaneously and in a collaborative manner. The multitask learning problem is relevant in several machine learning formulations and has been studied in the machine learning community in several contexts. For example, the problem finds applications in web page categorization~\cite{chen2009mtl}, web-search ranking~\cite{Chapelle2011mtl}, \cblue{and} disease progression modeling~\cite{zhou2011mtl}, among other areas. Clearly, this concept \cblue{is also relevant in the context of} distributed estimation  and adaptation over networks. Initial investigations along these lines for the traditional diffusion strategy appear in~\cite{Zhao2012,chen2013performance}.  In this article, we consider the general situation where there are connected clusters of nodes, and each cluster has a parameter vector to estimate. The estimation still needs to be performed cooperatively across the network because the data across the clusters may be correlated and, therefore, cooperation across clusters can be beneficial. Obviously, a limit case of this problem is the situation where all clusters are of equal size one, that is, each node has its own parameter vector to estimate but shares information with its neighbors. Another limit case is when the size of the cluster agrees with the size  of the network in which case all nodes have the same parameter vector to estimate. The aim of this paper is to derive diffusion strategies that are able to solve \cblue{this general} multitask estimation problem, and to analyze their performance in terms of mean-square error and convergence rate. Simulations are also conducted to illustrate the theoretical analysis, and to apply the algorithms to three useful applications involving spectral sensing, target localization, and hyperspectral data unmixing.

This paper is organized as follows. Section \ref{sec:intro-multitask} formulates the distributed estimation problem for multitask learning.  Section III presents a relaxation strategy for optimizing local cost functions over the network. Section IV derives a stochastic gradient algorithm for distributed adaptive learning in a multitask-oriented environment. Section V analyzes the theoretical performance of the proposed algorithm, in the mean and mean-square-error sense. In Section VI, experiments  and applications are presented to illustrate the performance of the approach. Section VII concludes this paper and gives perspectives on future work.

%-------------------------

\section{Network models and multitask learning}
\label{sec:intro-multitask}

Before starting our presentation, we provide a summary of \cblue{some of the main} symbols used in the article. Other symbols will be defined in the context where they are used:
\begin{tabbing}\small
 \hspace{0.5cm}  \=$x$   \hspace{2cm}	\=Normal font denotes scalars.  \\
\>$\bx$                        				\>Boldface small letters denote vectors. All vectors are column vectors.  \\
\>$\bR$                       				\>Boldface capital letters denote matrices.  \\
\>$(\cdot)^\top$         				\>Matrix \cred{transpose}.  \\
\>$\bI_N$                   				\>Identity matrix of size $N\times N$.  \\
\>$\N{k}$                     				\>The index set of nodes that are in the neighborhood of node $k$, including $k$.  \\
\>$\N{k}^-$                 				\>The index set of nodes that are in the neighborhood of node $k$, excluding $k$.  \\
\>$\C_i$                       				\>Cluster $i$, i.e., index set of nodes in the $i$-th cluster.  \\
\>$\C(k)$                     				\>The cluster to which node $k$ belongs, i.e., $\C(k) = \{\C_i : k \in \C_i\}$. \\
\>$J(\cdot)$,\, $\overline{J}(\cdot)$    	\> Cost functions without/with  regularization.   \\
\>$\bw^\star$,\, $\bw^o$                   		\> Optimum parameter vectors without/with regularization.
\end{tabbing}

\begin{figure*}[!t]
	\subfigure[Single-task network]{
 	\label{fig:UniqueW}
   	\begin{minipage}[c]{.3\linewidth}
   		\centering
      		\includegraphics[trim = 0mm 0mm 0mm 0mm, clip, scale=0.45]{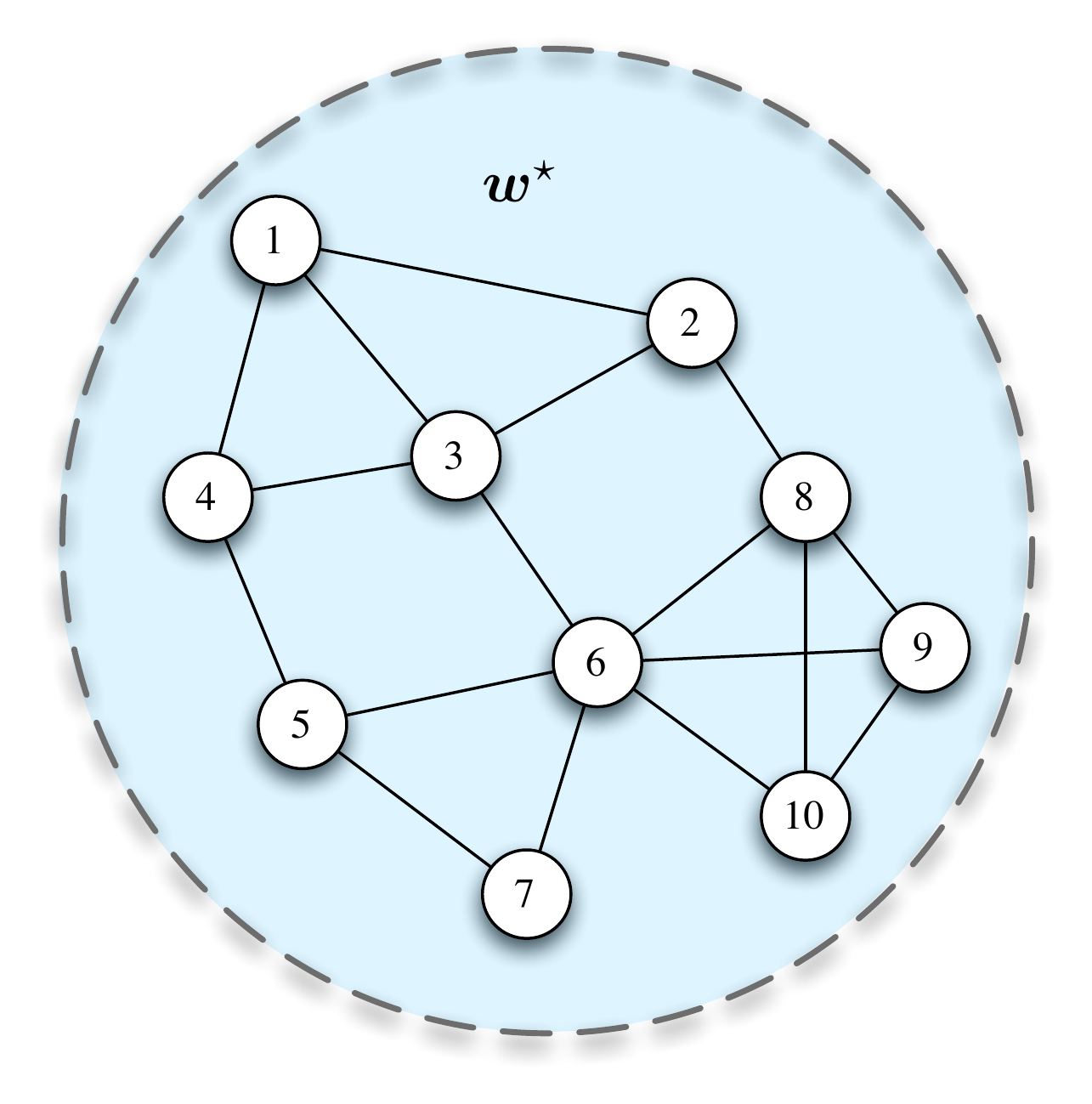}
   	\end{minipage}} \hfill
	\subfigure[Multitask network]{
 	\label{fig:DiffW}
   		\begin{minipage}[c]{.3\linewidth}
   		\centering
      		\includegraphics[scale=0.45]{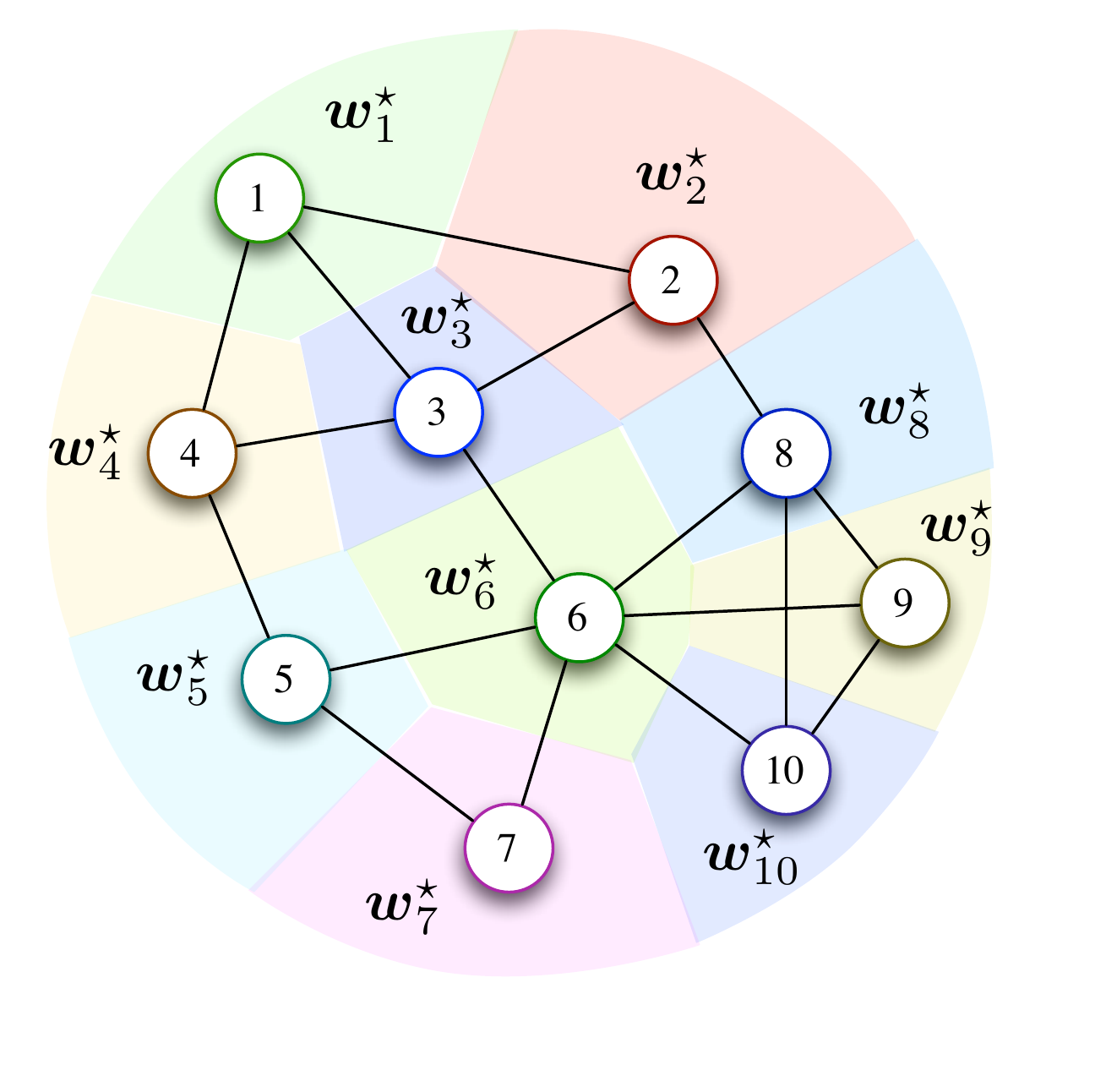}
   	\end{minipage}} \hfill
	\subfigure[Clustered multitask network]{
 	\label{fig:ClassW}
   		\begin{minipage}[c]{.3\linewidth}
   		\centering
      		\includegraphics[trim = 0mm 0mm 0mm 0mm, clip, scale=0.45]{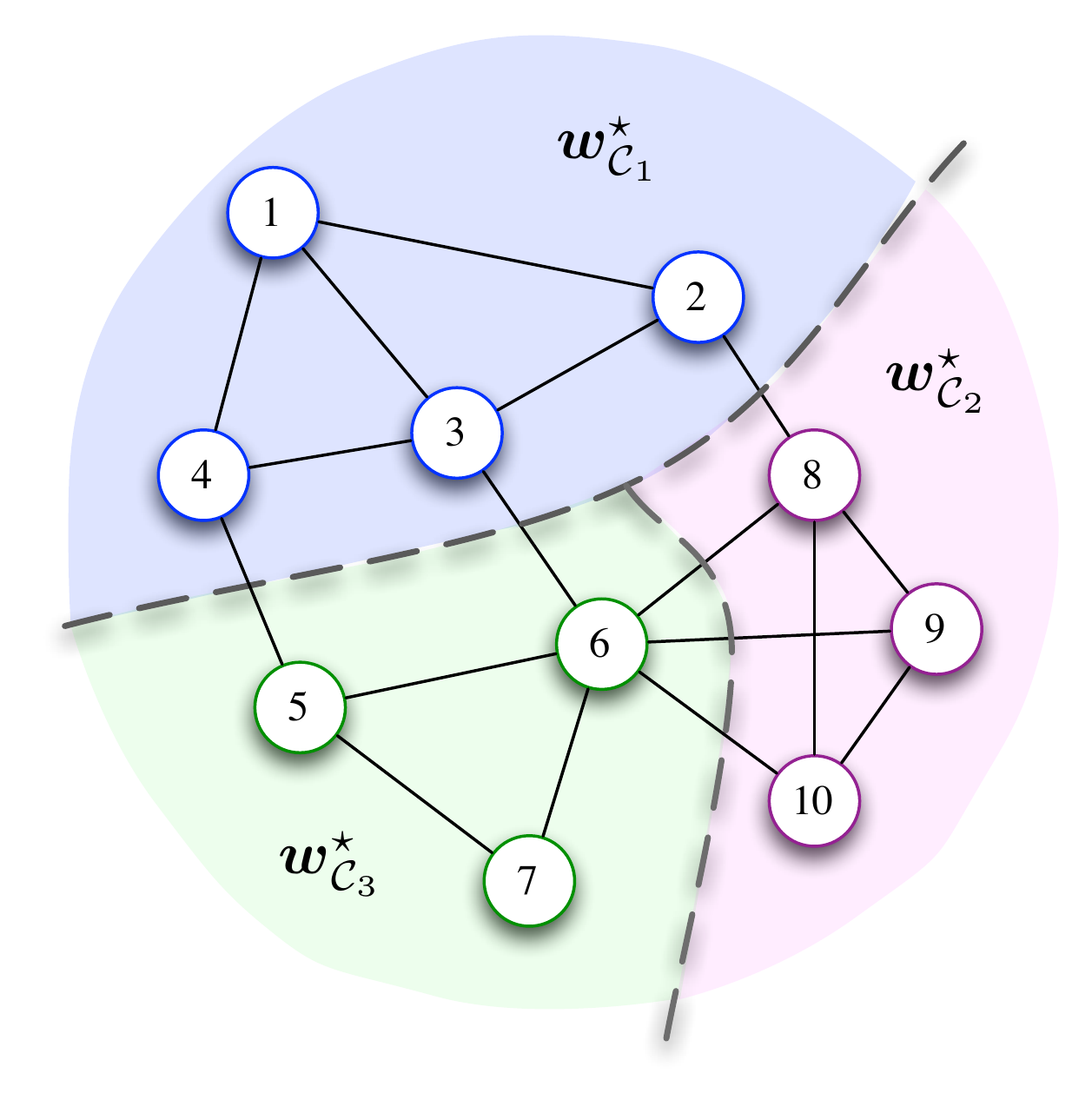}
   	\end{minipage}}
	\caption{Three types of networks. The single-task and multitask networks can be viewed as special cases of the clustered multitask network.}
	\label{fig:Network_Type}
\end{figure*}

We consider a connected network consisting of $N$ nodes. The problem is to estimate an $L\times 1$ unknown vector at each node $k$ from collected measurements. Node $k$ has access to temporal measurement sequences $\{d_k(n), \bx_k(n)\}$, with $d_k(n)$ denoting a scalar zero-mean reference signal, and $\bx_k(n)$ denoting an $L\times 1$ regression vector with a positive-definite \cblue{covariance} matrix, $ \bR_{x,k}=E\{\bx_k(n)\bx_k^\top(n)\} >0$. The data at node $k$ are assumed to be related via the linear regression model:
	\begin{equation}
		\label{eq:datamodel}
		d_k(n)=\bx_k^\top(n)\, \bw_k^\star + z_k(n)
	\end{equation}
where $\bw_{k}^{\star}$ is an unknown parameter vector at node $k$, and   $z_k(n)$ is a zero-mean i.i.d. noise that is independent of any other signal and has variance $\sigma_{z,k}^2$. Considering the number of parameter vectors to estimate, which we shall refer to as the number of tasks, the distributed learning problem can be single-task or multitask oriented. We therefore distinguish among the following three types of networks, as illustrated by Figure \ref{fig:Network_Type}, depending on how the parameter vectors $\bw_k^\star$ across the nodes are related:
\begin{itemize}
\item{\underline{Single-task networks}:} All nodes have to estimate the same parameter vector $\bw^\star$. That is, in this case we have that
	\begin{equation}
		\bw_k^\star = \bw^ \star, \quad \forall k\in\{1, \dots, N\}
	\end{equation}
\item{\underline{Multitask networks}:} Each node $k$ has to determine its own optimum parameter vector, $\bw_k^\star$. However, it is assumed that similarities and relationships exist among the parameters of neighboring nodes, which we denote by writing
	\begin{equation}
		\bw^\star_k \sim \bw^\star_\ell    \quad \text{if } \ell \in\N{k}
	\end{equation}
The sign $\sim$ represents a similarity relationship in some sense, and its meaning will become clear soon once we introduce expressions~\eqref{eq:MSEl2} and~\eqref{eq:P2} further ahead. Within the area of machine learning, the relation between tasks can be promoted in several ways, e.g., through mean regularization~\cite{Evgeniou2004}, low rank regularization~\cite{Ji2009}, or clustered regularization~\cite{Zhou2011}. We note that a number of application problems can be addressed using this model. For instance, consider an image sensor array and the problem of image restoration. In this case, links in Figure~\ref{fig:DiffW} can represent neighboring relationships between adjacent pixels. We will consider this application in greater detail in the \cblue{simulation section}.
\item{\underline{Clustered multitask networks}:} Nodes are grouped into $Q$ clusters, and there is one task per cluster. The optimum parameter vectors are only constrained to be equal within each cluster, but similarities between neighboring clusters are allowed to exist, namely,
	\begin{eqnarray}
			& \bw_k^\star = \bw^\star_{\C_q}, 		& \text{whenever }   k\in \C_q   \\
                        	& \bw^ \star_{\C_p} \sim \bw^\star_{\C_q},	& \text{if } \C_p, \, \C_q \text{ are connected}
	\end{eqnarray}
where $p$ and $q$ denote two cluster indexes. We say that two clusters ${\C}_p$ and ${\C}_q$ are connected if there exists at least one edge linking a node from one cluster to a node in the other cluster. 
\end{itemize}
One can observe that the single-task and multitask networks are particular cases of the clustered multitask network. In the case where all the nodes are clustered together, the clustered multitask network reduces to the single-task network. On the other hand, in the case where each cluster only involves one node, the clustered multitask network becomes a multitask network. Building on the literature on  diffusion strategies for single-task networks, we shall now generalize its use and analysis for distributed learning over clustered multitask networks. The results will be applicable to multitask networks by setting the number of clusters equal to the number of nodes.

%-------------------------

\section{Problem formulation}

\subsection{Global cost function and optimization}

Clustered multitask networks require that nodes that are grouped in the same cluster estimate the same coefficient vector. Thus, consider  the cluster $\C(k)$ to which node $k$ belongs. A local cost function, $J_k(\bw_{\C(k)})$, is associated with node $k$ and it  is assumed to be strongly convex and second-order differentiable, an example of which is the mean-square error criterion defined by
\begin{equation}
	\label{eq:JkMSE}
	J_k(\bw_{\C(k)}) = E\left\{\big|d_k(n) - \bx_k^\top(n)\,\bw_{\C(k)}\big|^2\right\}.
\end{equation}
In order to promote similarities among adjacent clusters, appropriate regularization can be used. For this purpose, we introduce the squared Euclidean distance as a possible regularizer, namely,
\begin{equation}
       \label{eq:l2dist}
	\Delta(\bw_{\C(k)}, \bw_{\C(\ell)}) = \|\bw_{\C(k)} - \bw_{\C(\ell)}\|^2.
\end{equation}
Combining \eqref{eq:JkMSE} and \eqref{eq:l2dist} yields the following regularized problem $\mathcal{P}_1$ at the level of the entire network:
\begin{equation}
         \label{eq:MSEl2}
         (\mathcal{P}_1)  \qquad
                 \overline{J^{\text{glob}}}(\bw_{\C_1}, \dots, \bw_{\C_Q}) = \sum_{k=1}^{N}
                 E\left\{\big|d_k(n) - \bx_k^\top(n)\,\bw_{\C(k)}\big|^2\right\}  + {\eta} \,
                 \sum_{k=1}^{N} \sum_{\ell\in\N{k} \backslash \C(k)} \rho_{k\ell} \,\|\bw_{\C(k)} - \bw_{\C(\ell)}\|^2,
\end{equation}
where $\bw_{\C_i}$ is the parameter vector associated with cluster $\C_i$ and $\eta>0$. The second term on the right-hand-side of expression~\eqref{eq:MSEl2} promotes similarities between the $\bw_{\C_i}$ of neighboring clusters, with strength parameter $\eta$.

Observe from the right-most term in~\eqref{eq:MSEl2} that the regularization strength between two clusters is directly related to the number of edges that connect them. The non-negative coefficients $\rho_{k\ell}$ aim at adjusting the regularization strength but they do not necessarily enforce symmetry. That is, we do not require $\rho_{k \ell} = \rho_{\ell k}$ even though the regularization term $\|\bw_{\C(k)}-\bw_{\C(\ell)}\|^2$ is symmetric with respect to the weight vectors $\bw_{\C(k)}$ and $\bw_{\C(\ell)}$; this term will be weighted by the sum $\rho_{k\ell}+\rho_{\ell k}$ due to the summation over the $N$ nodes. Consequently, problem formulation $\mathcal{P}_1$ inevitably leads to symmetric regularization despite the fact that $\rho_{k\ell}\neq\rho_{\ell k}$. However, we would like the design problem to benefit from the additional flexibility that is afforded by the use of asymmetric regularization coefficients. This is because asymmetry  \cblue{allows} clusters to \cblue{scale} their desire for closer similarity with their neighbors differently. For example, asymmetric regularization would allow  cluster $\C_k$ to promote similarities with cluster $\C_{\ell}$ while cluster $\C_{\ell}$ may be less inclined towards promoting similarities with $\C_k$. In order to exploit this flexibility more fully, we consider an alternative problem formulation $\mathcal{P}_2$ defined in terms  of $Q$ Nash equilibrium problems as follows:
\begin{equation}
	\label{eq:P2}
	\hspace{-5mm}(\mathcal{P}_2) \hspace{5mm} \left\{
	\begin{split}
		&\min_{\bw_{\C_i}}\overline{J_{\C_i}}(\bw_{\C_i},\bw_{-\C_i}) \qquad \text{for } i = 1, \dots, Q \\
		&\text{with }\; \overline{J_{\C_i}}(\bw_{\C_i},\bw_{-\C_i}) = \sum_{k\in\C_i}  E\left\{\big|d_k(n) - \bx_k^\top(n)\,\bw_{\C(k)}\big|^2\right\} 
		+ \eta\,\sum_{k\in\C_i}\sum_{\ell\in\N{k}\backslash\C_i} \rho_{k\ell}  \,\|\bw_{\C(k)} - \bw_{\C(\ell)}\|^2
	\end{split} \right.
\end{equation}
where each cluster $\C_i$ estimates $\bw_{\C_i}$ by minimizing $\overline{J_{\C_i}}(\bw_{\C_i},\bw_{-\C_i})$. Note that we have kept the notation $\bw_{\C(k)}$ to make the role of the regularization term clearer, even though in formulation~\eqref{eq:P2} we have $\bw_{\C(k)}=\bw_{\C_i}$ for all $k$ in $\C_i$. In~\eqref{eq:P2}, the notation $\bw_{-\C_i}$ denotes the collection of weight vectors estimated by the other clusters, i.e., $\bw_{-\C_i} = \{\bw_{\C_k}:k=1,\ldots,Q\}-\{\bw_{\C_i}\}$.

The Nash equilibrium of $\mathcal{P}_2$ satisfies the condition~\cite{Dynamic1995}:
\begin{equation}
	\bw_{\C_i}^o = \arg\min_{\bw_{\C_i}} \overline{J_{\C_i}}(\bw_{\C_i},\bw^o_{-\C_i})
\end{equation}
for $i = 1, \dots, Q$, where the notation $\bw^o_{-\C_i}$ denotes the collection of the Nash equilibria by the other clusters. Problem $\mathcal{P}_2$ has the following properties:
\begin{enumerate}
	\item An equilibrium exists for $\mathcal{P}_2$ since $\overline{J_{\C_i}}(\bw_{\C_i},\bw_{-\C_i})$ is convex with respect to $\bw_{\C_i}$ for all $i$.
	\item The equilibrium for $\mathcal{P}_2$ is unique since $\{\overline{J_{\C_i}}(\bw_{\C_i},\bw_{-\C_i})\}_{i=1}^Q$ satisfies the diagonal strict
	convexity property.\footnote{Let $\bg(\bw,\bzeta)=[\zeta_i\nabla_{\bw_{\C_i}} \overline{J_{\C_i}}(\bw_{\C_i},\bw_{-\C_i})]_{i=1}^Q$ arranged as a row
	vector with $\zeta_i>0$. The cost functions $\{\overline{J_{\C_i}}(\bw_{\C_i},\bw_{-\C_i})\}_{i=1}^Q$ satisfy the diagonal strict convexity property 
	if $\bg(\bw,\bzeta)$ is strictly decreasing in $\bw$ for some positive vector $\bzeta$, that is, $(\bg(\hat\bw,\bzeta)-\bg(\bw,\bzeta))^\top(\hat\bw-\bw)<0$
	for all nonequal $\bw$, $\hat\bw$.}
	\item Problems $\mathcal{P}_1$ and $\mathcal{P}_2$ have the same solution by setting the value of $\rho_{k\ell}$ in $\mathcal{P}_2$ to  that of
	$\rho_{k\ell}+\rho_{\ell k}$ from $\mathcal{P}_1$.
\end{enumerate}
Properties 1) and 2) can be checked via Theorems 1 and 2 in~\cite{rosen1965}. Property 3) can be verified by the optimality conditions for the two problems.

Problem $\mathcal{P}_1$ can be solved either analytically in closed form or  iteratively by using a steepest-descent algorithm. Unfortunately, there is no analytical expression for general Nash equilibrium problems. We estimate the equilibrium of problem $\mathcal{P}_2$ iteratively by the fixed point of the best response iteration~\cite{Dynamic1995}, that is,
\begin{equation}
	\label{eq:br}
	\bw_{\C_i}(n+1) = \arg\min_{\bw_{\C_i}} \overline{J_{\C_i}}(\bw_{\C_i},\bw_{-\C_i}(n))
\end{equation}
for $i = 1, \dots, Q$, and leads to the solution of~\eqref{eq:P2}. Since the equilibrium is unique and the cost function for each cluster is convex, the solution of~\eqref{eq:P2} can also be \cblue{approached} by means of a steepest-descent iteration as follows:
\begin{equation}
	\label{eq:gd}
	\bw_{\C_i}(n+1) = \bw_{\C_i}(n) - \mu \, \nabla_{\bw_{\C_i}} \overline{J_{\C_i}}(\bw_{\C_i}(n),\bw_{-\C_i}(n))
\end{equation}
for $i = 1, \dots, Q$, with $\nabla_{\bw_{\C_i}}$ denoting the gradient operation with respect to $\bw_{\C_i}$, and $\mu$ a positive step-size. We have
\begin{equation}
	\label{eq:grd}
	\nabla_{\bw_{\C_i}} \overline{J_{\C_i}}(\bw_{\C_i},\bw_{-\C_i}(n)) \propto \sum_{k\in\C_i}(\bR_{x,k}\bw_{\C_i}
		- \cb{p}_{xd,k}) +  \eta\, \sum_{k\in\C_i}\sum_{\ell\in\N{k}\backslash\C_i} \rho_{k\ell} \big( \bw_{\C_i} -  \bw_{\C(\ell)}(n)\big).
 \end{equation}
 where $\bp_{xd,k} = E\{\bx_k(n)d_k(n)\}$ is the input-output cross-correlation vector between $\bx_k(n)$ and $d_k(n)$ at node~$k$. If some additional constraints are imposed on the parameters to estimate, the gradient update relation can be modified using methods such as projection~\cite{Theodoridis2011adap} or fixed point iteration techniques~\cite{chen2011nnlms}. In the body of the paper, we  focus on the unconstrained case during the algorithm derivation and its analysis. However, a constrained problem will be presented in the simulation section. \cblue{Since} $\mathcal{P}_1$ is equivalent to $\mathcal{P}_2$ with proper setting of the weights $\rho_{k\ell}$, we shall now derive a distributed algorithm for solving problem $\mathcal{P}_2$. In this paper, we shall consider normalized weights that satisfy
\begin{equation}
	\label{eq:rho}
	\sum_{\ell=1}^N \rho_{k\ell} = 1,\quad \text{and}\quad \rho_{k\ell} = 0 \; \text{ if }\, \ell\notin\N{k}\backslash \C(k).
\end{equation}

%----------------------- subsection : local cost -----------------------
\subsection{Local cost decomposition and problem relaxation}

The solution method~\eqref{eq:gd} using~\eqref{eq:grd} requires that every node in the network should have access to the statistical moments $\bR_{x,k}$ and $\bp_{xd,k}$ over its cluster. There are two problems with this scenario. First, nodes can only be assumed to have access to information from their immediate neighborhood and the cluster of every node $k$ may include nodes that are not direct neighbors of $k$.  Second, nodes rarely have access to the moments $\{\bR_{x,d}, \bp_{xd,k}\}$; instead, they have access to data generated from distributions with these moments. Therefore, more is needed to enable a distributed solution that relies solely on local interactions within neighborhoods and that relies on measured data as opposed to statistical moments.  To derive a distributed algorithm, we follow the approach of~\cite{Cattivelli2010diff,Sayed2013intr}. The first step in this approach is to show how to express the cost~\eqref{eq:P2} in terms of other local costs that only depend on data from neighborhoods.

Thus, let us introduce an $N\times N$ right stochastic matrix $\bC$ with nonnegative entries $c_{\ell k}$ such that
\begin{equation}
	\sum_{k=1}^N c_{\ell k} = 1, \quad \text{and} \quad c_{\ell k} =0 \,\text{ if }\, k\notin \cp{N}_{\ell} \cap \C(\ell).
\end{equation}
With these coefficients, we associate a local cost function of the following form with each node $k$~\cite{Sayed2013intr}:
\begin{equation}
	\label{eq:Jloc}
	J_k^{\text{loc}}(\bw_{\C(k)}) =
	\sum_{\ell\in\N{k}\cap\C(k)}  c_{\ell k} \, E\left\{\big|d_\ell(n) - \bx_\ell^\top(n)\,\bw_{\C(k)}\big|^2\right\}.
\end{equation}
One important distinction from the local cost defined in~\cite{Sayed2013intr} is that in~\cite{Sayed2013intr} the summation in~\eqref{eq:Jloc} is defined over the entire neighborhood of node $k$, i.e., for all $\ell \in \N{k}$. Here we are excluding those neighbors of $k$ that do not belong to its cluster. This is because these particular neighbors will be pursuing a different parameter vector than node $k$. Furthermore, we note in~\eqref{eq:Jloc} that $\bw_{\C(k)} = \bw_{\C(\ell)}$  because $\ell \in \C(k)$. To make the notation simpler, we shall write $\bw_k$ instead of $\bw_{\C(k)}$. A consequence of this notation is that $\bw_k=\bw_\ell$ for all $\ell\in\C(k)$.
Incorporating the estimates of the neighboring clusters, we modify~\eqref{eq:Jloc} to associate a regularized local cost function with node $k$ of the following form
\begin{equation}
	\label{eq:Jlocr}
	\overline{J_k^{\text{loc}}}(\bw_k) = \sum_{\ell\in\N{k}\cap\C(k)}  c_{\ell k} 
	\,E\left\{\big|d_\ell(n) - \bx_\ell^\top(n)\,\bw_k\big|^2\right\} + \eta \sum_{\ell\in\N{k}\backslash\C(k)} \rho_{k\ell} \|\bw_k - \bw_\ell\|^2.
\end{equation}
\cmag{Observe that this local cost is now solely defined in terms of information that is available to node $k$ from its neighbors}. Using this regularized local cost function, it can be verified that the global cost function for cluster $\C_i$ in~\eqref{eq:P2} can be now expressed as
\begin{align}
	\overline{J_{\C_i}}(\bw_{\C_i},\bw_{-\C_i}) &=   \sum_{k\in\C_i} \Big(\sum_{\ell\in \C(k)}  c_{\ell k}   E\left\{\big|d_\ell(n) - \bx_\ell^\top(n)\,\bw_{\C(k)}\big|^2\right\} 
		+ \eta\,\sum_{\ell\in\N{k}\backslash\C_i} \rho_{k\ell}  \,\|\bw_{\C(k)} - \bw_{\C(\ell)}\|^2 \Big) \nonumber\\
	 &= \sum_{k\in\C_i} \overline{J_k^{\text{loc}}}(\bw_k)  \nonumber\\
	 &=  \overline{J_k^{\text{loc}}}(\bw_k) + \sum_{\ell\in\C(k) \backslash k } \overline{J_\ell^{\text{loc}}}(\bw_\ell)  \label{eq:JclusterSum}
\end{align}
Let $\bw_k^o$ denote the minimizer of the local cost function~\eqref{eq:Jlocr}, given $\bw_{\ell}$ for all $\ell\in\N{k}\backslash\C(k)$. A completion-of-squares argument shows that each $\overline{J_k^{\text{loc}}}(\bw_k)$ can be expressed as
\begin{equation}
	\label{eq:local.reg.eqv}
	\overline{J_k^{\text{loc}}}(\bw_k) =  \overline{J_k^{\text{loc}}}(\bw_k^o) 
	+ \|\bw_k - \bw_k^o\|^2_ {\overline{\bR}_{k}}
\end{equation}
where 
\begin{equation}
         \overline{\bR}_{k}=\sum_{\ell\in\N{k}\cap\C(k)} c_{\ell k}\,\bR_{x, \ell} + \eta\sum_{\ell\in\N{k}\backslash\C(k)} \rho_{k\ell}\bI_L.
 \end{equation}
Substituting equation~\eqref{eq:local.reg.eqv} into the second term on the right-hand-side of~\eqref{eq:JclusterSum}, and discarding the terms $\{\overline{J_\ell^{\text{loc}}}(\bw_\ell^o)\}$because they are independent of the optimization variables in the cluster, we can consider the following equivalent cost function for  cluster $\C(k)$ at node $k$:
\begin{equation}
	\label{eq:Jglob.eqv}
         \overline{J_{\C(k)}}(\bw_{k}) \triangleq \overline{J_k^{\text{loc}}}(\bw_k)+\sum_{\ell \in\C(k)\backslash k}  \|\bw_\ell - \bw^o_\ell\|^2_{\overline{\bR}_{\ell}}
 \end{equation}
where it holds that $\bw_k = \bw_\ell$  because $\ell \in \C(k)$. Note that we have omitted $\bw_{-k}$ in the notation for $\overline{J_{\C(k)}}(\bw_{k})$ for the sake of brevity. Therefore, minimizing~\eqref{eq:Jglob.eqv}  is equivalent to minimizing the original cost~\eqref{eq:JclusterSum} or~\eqref{eq:P2} over $\bw_k$. However the second term~\eqref{eq:Jglob.eqv}  still requires information from nodes $\ell$ that may not be in the direct neighborhood of node $k$ even though they belong to the same cluster. In order to avoid access to information via multi-hop, we can relax the cost function~\eqref{eq:Jglob.eqv} at node $k$ by considering only information originating from its neighbors. This can be achieved by replacing the range of the index over which the summation in~\eqref{eq:Jglob.eqv} is computed as follows:
\begin{equation}
	\label{eq:Jglob.eqv.relax}
		\overline{J_{\C(k)}}'(\bw_k) = \overline{J_k^{\text{loc}}}(\bw_k)+\sum_{\ell\in \N{k}^-\cap\C(k)}   \|\bw_k - \bw^o_\ell\|^2_{\overline{\bR}_{\ell}}.
 \end{equation}
Usually, especially in the context of adaptive learning in a non-stationary environment, the weighting matrices $\overline{\bR_{\ell}}$ are unavailable since the covariance matrices $\bR_{x,\ell}$ at each node may not be known beforehand. Following an argument based on the Rayleigh-Ritz characterization of eigenvalues, it was explained in~\cite{Sayed2013intr} that a useful strategy is to replace each matrix $\overline{\bR}_{\ell}$ by a weighted multiple of the identity matrix, say, as:
\begin{equation}
	\|\bw_k - \bw^o_\ell\|^2_{\overline{\bR}_{\ell}} \approx {b_{\ell k}} \, \|\bw_k - \bw^o_\ell\|^2
\end{equation}
for some nonnegative coefficients $b_{\ell k}$ that can possibly depend on the node $k$. As shown later, these coefficients will be incorporated into a left stochastic matrix to be defined and, therefore, the designer does not need to worry about the selection of the $b_{\ell k}$ at this stage. Based on the arguments presented so far, and using~\eqref{eq:Jlocr}, the global cost function \eqref{eq:Jglob.eqv.relax} can then be relaxed to the following form:
\begin{equation}
	\label{eq:Jglob.eqv.relax2}
         \begin{split}
         \overline{J_{\C(k)}}''(\bw_k) = \sum_{\ell\in\N{k}\cap\C(k)} c_{\ell k} &\, E\left\{\big|d_\ell(n) - \bx_\ell^\top(n)\,\bw_k\big|^2\right\} \\
         	&+ \eta \sum_{\ell\in\N{k}\backslash\C(k)} \rho_{k\ell}\,\|\bw_k - \bw_\ell\|^2 + \sum_{\ell\in \N{k}^-\cap\C(k)} {b_{\ell k}}\|\bw_k - \bw^o_\ell\|^2.
         \end{split}
 \end{equation}
Observe that the two last sums on the right-hand-side of~\eqref{eq:Jglob.eqv.relax2} divide the neighbors of node $k$ into two exclusive sets: those that belong to its cluster (last sum) and those that do not belong to its cluster (second term). In summary, the argument so far enabled us to replace the cost~\eqref{eq:P2} by the alternative cost~\eqref{eq:Jglob.eqv.relax2} that depends only on data within the neighborhood of node $k$. We can now proceed to use~\eqref{eq:Jglob.eqv.relax2} to derive distributed strategies. Subsequently, we study the stability and mean-square performance of the resulting strategies and show that they are able to perform well despite the approximation introduced in steps.

\section{Stochastic approximation algorithms}

To begin with, a steepest-descent iteration can be applied by each node $k$ to minimize the cost function \eqref{eq:Jglob.eqv.relax2}. Let $\bw_k(n)$ denote the estimate for $\bw_k$ at iteration $n$. Using a constant step-size $\mu$ for each node, the update relation would take the following form:
\begin{equation}
	\label{eq:local.update}
	\begin{split}
		\bw_k(n+1) = \bw_k(n) &-\mu \hspace{-3mm} \sum_{\ell\in\N{k}\cap\C(k)} \hspace{-3mm} c_{\ell k}\left(\bR_{x,\ell}\bw_k(n) - \cb{p}_{xd,k}\right)
		- \mu\eta\hspace{-3mm}\sum_{\ell\in\N{k} \backslash \C(k)} \hspace{-3mm} \rho_{k\ell}\, \left(\bw_{k}(n)-\bw_{\ell}(n)\right) \\
		& - \mu\hspace{-3mm}\sum_{\ell\in{\N{k}^-\cap\C(k)}}\hspace{-3mm} b_{\ell k}\,\left(\bw_k(n)- \bw^o_\ell\right)
	\end{split}
\end{equation}
Among other possible forms,  expression~\eqref{eq:local.update} can be evaluated in two successive update steps
\begin{align}
	&\bpsi_k(n+1)	= \bw_k(n) - \mu\left( \sum_{\ell\in\N{k}\cap\C(k)} \hspace{-3mm} c_{\ell k}\left(\bR_{x,\ell}\bw_k(n) - \cb{p}_{xd,k}\right)
		+\eta\hspace{-3mm}\sum_{\ell\in\N{k} \backslash \C(k)} \hspace{-3mm} \rho_{k\ell}\, (\bw_{k}(n)-\bw_{\ell}(n))\right) \label{eq:step1}\\
	&\bw_k(n+1)	= \bpsi_k(n+1) + \mu\hspace{-3mm}\sum_{\ell\in{\N{k}^-\cap\C(k)}}
		\hspace{-3mm} b_{\ell k}\,\left(\bw^o_\ell-\bw_k(n)\right)\label{eq:step2}
\end{align}
Following the same line of reasoning from~\cite{Sayed2013intr} in the single-task case, and extending the argument to apply to clusters, we use $\bpsi_\ell(n+1)$ as a local estimate for  $\bw^o_\ell$ in~\eqref{eq:step2} since the latter is unavailable and $\bpsi_\ell(n+1)$ is an intermediate estimate for it that is available at node $\ell$ at time $n+1$. In addition, again in step \eqref{eq:step2}, we replace $\bw_k(n)$ by $\bpsi_k(n+1)$ since it is a better estimate obtained by incorporating information from the neighbors according to~\eqref{eq:step1}. Step \eqref{eq:step2} then becomes
\begin{equation}
         \label{eq:wpsi}
	\bw_k(n+1) = \left(1-  \mu \hspace{-3mm}  \sum_{\ell\in{\N{k}^-\cap\C(k)}}\hspace{-3mm} b_{\ell k} \right) \bpsi_k(n+1) 
	+ \mu \hspace{-3mm} \sum_{\ell\in{\N{k}^-\cap\C(k)}} \hspace{-3mm} b_{\ell k}\, \bpsi_\ell(n+1).
\end{equation}
The coefficients in~\eqref{eq:wpsi} can be redefined as:
\begin{equation}
	\begin{split}
		a_{kk} 	& \triangleq 1-  \mu   \sum_{\ell\in{\N{k}^-\cap\C(k)}} b_{\ell k}      \\
		a_{\ell k} 	& \triangleq \mu\, b_{\ell k},  \quad \ell\in\N{k}^-\cap\C(k)   \\
		a_{\ell k } 	& \triangleq 0, \quad  \ell\notin\N{k}\cap\C(k)
	\end{split}
\end{equation}
It can be observed that the entries $\{a_{\ell k}\}$ are nonnegative for all $\ell$ and $k$ (including $a_{kk}$) for sufficiently small step-size. Moreover, the matrix $\bA$ with ($\ell,k$)-th entry $a_{\ell k}$ is a left-stochastic matrix, which means that the sum of each of its columns is equal to one. With this notation, we obtain the following adapt-then-combine (ATC) diffusion strategy for solving problem~\eqref{eq:P2} in a distributed manner:
%\medskip
\begin{equation}
	\label{eq:ATC}
	\begin{split}
	&\bpsi_k(n+1)	= \bw_k(n) - \mu\,\left(\sum_{\ell\in\N{k}\cap\C(k)} \hspace{-3mm} c_{\ell k}\left(\bR_{x,\ell}\bw_k(n) - \cb{p}_{xd,k}\right)
		+\eta\hspace{-3mm}\sum_{\ell\in\N{k} \backslash \C(k)} \hspace{-3mm} \rho_{k\ell}\, (\bw_{k}(n)-\bw_{\ell}(n))\right)\\
	&\bw_k(n+1) = \sum_{\ell\in\N{k}\cap\C(k)} \hspace{-3mm} a_{\ell k} \, \bpsi_k(n+1)
	\end{split}
\end{equation}
At each instant $n+1$, node $k$ updates the intermediate value $\bpsi_k(n+1)$ with a local steepest descent iteration. This step involves a regularization term in the case where the set of inter-cluster neighbors of node $k$ is not empty. Next, an aggregation step is performed where node $k$ combines its intermediate value $\bpsi_k(n+1)$ with the intermediate values $\bpsi_\ell(n+1)$ from its cluster neighbors. It is also possible to arrive at a combine-then-adapt (CTA) diffusion strategy where the aggregation step is performed prior to the adaptation step~\cite{Sayed2013intr}. In what follows, it is sufficient to  focus on the ATC  strategy to illustrate the main results. Employing instantaneous approximations for the required signal moments in~\eqref{eq:ATC}, we arrive at the desired  diffusion strategy for clustered multitask learning described in Algorithm \ref{algo:diffLMS.hybmulti}  where the regularization factors $\rho_{k\ell}$ are chosen according to~\eqref{eq:rho}, and the coefficients $\{c_{\ell k}, a_{\ell k}\}$ are nonnegative scalars chosen at will by the designer to satisfy the following conditions:
\begin{align}
&a_{\ell k}\geq 0,\;\;\;\sum_{\ell\in{\cal N}_k\cap {\cal C}(k)} a_{\ell k}=1,\;\;\;a_{\ell k}=0\;\;\mbox{\rm for } \; \ell \notin {\cal N}_k\cap {\cal C}(k)\\
&c_{\ell k}\geq 0,\;\;\;\sum_{k\in{\cal N}_{\ell}\cap {\cal C}(\ell)} c_{\ell k}=1,\;\;\;c_{\ell k}=0\; \;\mbox{\rm for } \;k \notin {\cal N}_{\ell}\cap {\cal C}(\ell)
\end{align}
There are several ways to select these coefficients such as using the averaging rule or the Metropolis rule (see~\cite{Sayed2013intr} for a listing of these and other choices).
\begin{algorithm}
Start with $\bw_k(0) = 0$ for all $k$, and repeat:
\begin{equation}
	\label{eq:ATC_MSEl2}
     	\hspace{-3mm}\left\{
	\begin{split}
	&\bpsi_k(n+1)	= \bw_k(n) + \mu\left(\sum_{\ell\in\N{k}\cap\C(k)} \hspace{-3mm} c_{\ell k}\left(d_\ell(n)-\bx_\ell^\top(n)\bw_k(n)\right) \bx_\ell(n)
		+ \eta\hspace{-3mm}\sum_{\ell\in\N{k} \backslash \C(k)} \hspace{-3mm}\rho_{k\ell}\,(\bw_{\ell}(n) -  \bw_{k}(n))\right)\\
	&\bw_k(n+1) = \sum_{\ell\in\N{k}\cap\C(k)} \hspace{-3mm} a_{\ell k} \, \bpsi_k(n+1)
	\end{split} \right.
\end{equation}
\caption{Diffusion LMS for clustered multitask networks} \label{algo:diffLMS.hybmulti}
\end{algorithm}

In the case of a single-task network when there is a single cluster that consists of the entire set of nodes we get $\N{k}\cap\,\C(k) = \N{k}$ and $\N{k}\backslash\C(k) = \emptyset$ for all $k$, so that expression \eqref{eq:ATC_MSEl2} reduces to the  diffusion adaptation strategy~\cite{Sayed2013intr,Cattivelli2010diff} described in Algorithm~\ref{algo:diffLMS}.
\begin{algorithm}
Start with $\bw_k(0) = 0$ for all $k$, and repeat:
\begin{equation}
	\label{eq:ATC_diff}
    	\left\{
	\begin{split}
	&\bpsi_k(n+1) = \bw_k(n) + \mu \!\sum_{\ell\in\N{k}} \!\! c_{\ell k}\left(d_\ell(n)-\bx_\ell^\top(n)\bw_k(n)\right)\bx_\ell(n)  \\
	&\bw_k(n+1) = \sum_{\ell\in\N{k}}  a_{\ell k} \, \bpsi_k(n+1)
	\end{split} \right.
\end{equation}
\caption{Diffusion LMS for single-task networks~\cite{Lopes2008diff,Cattivelli2010diff}.}\label{algo:diffLMS}
\end{algorithm}

In the case of a multitask network where the size of each cluster is one, we have  $\N{k}\cap\C(k) = \{k\}$ and $\N{k}\backslash\C(k) = \N{k}^-$ for all $k$, the algorithm and~\eqref{eq:ATC_diff} degenerate into Algorithm~\ref{algo:diffLMS.multi}. Interestingly, this is the instantaneous gradient counterpart of equation~\eqref{eq:gd} for each node.
\begin{algorithm}[!ht]
Start with $\bw_k(0) = 0$ for all $k$, and repeat:
\begin{equation}
 	\label{eq:ATC_MSE_multi}
	\bw_k(n+1)	= \bw_k(n) + \mu \,\left(d_k(n)-\bx_k^\top(n)\bw_k(n)\right)\bx_k(n)
				+\eta\,\mu\sum_{\ell\in\N{k}^-}  \rho_{k\ell}\left(\bw_{\ell}(n)-\bw_{k}(n)\right)
\end{equation}
\caption{Diffusion LMS for multitask networks}\label{algo:diffLMS.multi}
\end{algorithm}

%\afterpage{\clearpage}
%\FloatBarrier
 %=================  Analysis ==========================

%\section{Behavior  analysis}
\section{Mean-Square Error Performance Analysis}

We now examine the stochastic behavior of the adaptive diffusion strategy~\eqref{eq:ATC_MSEl2}. In order to address this question, we collect information from across the network into block vectors and matrices. In particular, let us denote by $\bw(n)$, $\bw^\star$ and $\bpsi$ the block weight estimate vector, the block optimum weight vector and block intermediate weight estimate  vector, all of size $LN \times 1$, i.e.,
\begin{equation}
	\bw(n) = \left(\begin{array}{c} \bw_1(n) \\ \vdots \\  \bw_N(n)  	\end{array}\right),  \qquad\qquad
	\bw^\star = \left(\begin{array}{c} \bw^\star_1 \\ \vdots \\  \bw^\star_N		\end{array}\right),   \qquad\qquad
	\bpsi(n) = \left(\begin{array}{c} \bpsi_1(n) \\ \vdots \\  \bpsi_N(n)		\end{array}\right)
 \end{equation}
with $\bw^\star_k = \bw^\star_{\C(k)}$. The weight error vector for each node $k$ at iteration $n$ is defined by $\bv_k(n) = \bw_k(n) - \bw^\star_k$. The weight error vectors $\bv_k(n)$ are also stacked on top of each other  to get the block weight error vector defined as follows:
\begin{equation}
	\bv(n) = \left(\begin{array}{c} \bv_1(n) \\ \vdots \\  \bv_N(n)  \end{array}\right)
\end{equation}
To perform the theoretical analysis, we introduce the following independence assumption.

\medskip
\assumption (Independent regressors) The regression vectors $\bx_k(n)$ arise from a stationary random process that is temporally stationary, white and independent over space with $\bR_{x,k} = E\{\bx_k(n)\bx^\top_k(n)\}>0$.
\medskip

\noindent A direct consequence is that $\bx_k(n)$ is independent of $\bv_\ell(m)$ for all $\ell$ and $m\leq n$. Although not true in general, this assumption is commonly used for analyzing adaptive constructions because it allows to simplify the derivations without constraining the conclusions. \cblue{Moreover}, various analyses in the literature have already shown that performance results obtained under this assumption match well the actual \cblue{performance of adaptive algorithms} when the step-size is sufficiently small~\cite{sayed2008adaptive}.

\subsection{Mean error behavior analysis}

The estimation error that appears in the first equation of~\eqref{eq:ATC_MSEl2} can be rewritten as
\begin{equation}
	\begin{split}
		d_\ell(n)-\bx_\ell^\top(n)\bw_k(n) 	&= \bx_\ell^\top(n)\bw_k^\star + z_\ell(n) - \bx_\ell^\top(n)\bw_k(n) \\
									&= z_\ell(n) - \bx_\ell^\top(n)\bv_k(n)
	\end{split}
\end{equation}
because $\bw_\ell^\star=\bw_k^\star$ for all $\ell\in\N{k}\cap\C(k)$. Subtracting $\bw^\star_k$ from both sides of the first equation in~\eqref{eq:ATC_MSEl2}, and using the above relation, the update equation for the block weight error vector of $\bpsi_k(n+1)$ can be expressed as
\begin{equation}
	\label{eq:psi.v}
	\bpsi(n+1) - \bw^\star = \bv(n) - \mu\, \bH_x(n) \, \bv(n)  + \mu\,\bp_{zx}(n) - \mu\,\eta\, \bQ \, (\bv(n) +\bw^\star)
\end{equation}
where
\begin{equation}
          \label{eq:Q}
          \bQ=\bI_{LN}-\bP \otimes \bI_L,
\end{equation}
with $\otimes$ denoting the Kronecker product, and $\bP$ the $N\times N$ matrix with $(k,\ell)$-th entry $\rho_{k\ell}$. Moreover, the matrix $\bH_x(n)$ is block diagonal of size $LN\times LN$ defined as
 \begin{equation}
	\bH_{x}(n) = \text{diag}\Big\{\sum_{\ell\in\N{1}\cap \C(1)} c_{\ell 1 }\,\bx_\ell(n)\bx_\ell^\top(n),\,\dots,\!\!\!
	\sum_{\ell\in\N{N}\cap \C(N)} c_{\ell N }\, \bx_\ell(n)\bx_\ell^\top(n) \Big\},
 \end{equation}
 and  $\bp_{zx}(n)$ \cblue{is the following vector} of length $LN\times 1$:
 \begin{equation}
	\bp_{zx}(n) = \Big(\sum_{\ell\in\N{1}\cap \C(1)}  c_{\ell 1}\, \bx_\ell^\top(n)z_\ell(n),\,\dots,\!\!\!\sum_{\ell\in\N{N}\cap \C(N)}
	c_{\ell N}\, \bx_\ell^\top(n)z_\ell(n)\Big)^\top.
 \end{equation}
Let  ${\bA}_I = \bA \otimes \bI_L$. The second equation in~\eqref{eq:ATC_MSEl2} then allows us to write
\begin{equation}
	\bw(n+1) = \bA_I^\top\,\bpsi(n+1)
\end{equation}
Subtracting $\bw^\star$ from both sides of the above expression, and using equation~\eqref{eq:psi.v}, the update relation can be written in a single expression as follows
\begin{equation}
	\label{eq:v}
	\begin{split}
		\bv(n+1) 	%& = \bA_I^\top \left(\bv(n) - \mu\, \bH_x(n) \, \bv(n)  + \mu\,\bp_{zx}(n) - \mu\,\eta\, \bQ\, (\bv(n) +\bw^\star)\right) \\
				 = \bA_I^\top \left[ \bI_{LN} - \mu\, (\bH_x(n) + \eta \, \bQ)\right]\,\bv(n) + \mu\,\bA_I^\top\,\bp_{zx}(n)
					-\mu\,\eta\, \bA_I^\top\bQ\,\bw^\star
                 \end{split}
\end{equation}
Taking the  expectation of both sides, and using Assumption 1 we get
\begin{equation}
	\label{eq:Ev}
	E\{\bv(n+1)\} = \bA_I^\top \left[ \bI_{LN} - \mu\, (\bH_R + \eta \, \bQ)\right]\,E\{\bv(n)\} -  \mu\,\eta\,\bA_I^\top\bQ\, \bw^\star
\end{equation}
where 
\begin{equation}
              \bH_R  \triangleq  E\{\bH_x(n)\} = \text{diag}\left\{\bR_1,\,\dots,\bR_N\right\}
\end{equation}            
with 
\begin{equation}
              \bR_{k}=\sum_{\ell\in\N{k}\cap\C(k)} c_{\ell k}\,\bR_{x, \ell}.
\end{equation}

\theorem (Stability in the mean) Assume data model~\eqref{eq:datamodel} and Assumption 1 hold. Then, for any initial condition, the diffusion multitask strategy~\eqref{eq:ATC_MSEl2} asymptotically converges in the mean if the step-size is chosen to satisfy
\begin{equation}
          \label{eq:rhospr}
	\rho \left(\bA_I^\top \left[\bI_{LN} - \mu\, (\bH_R + \eta \, \bQ)\right]\right)<1
\end{equation}
where $\rho(\cdot)$ denotes the spectral radius of its matrix argument. A \emph{sufficient} condition for~\eqref{eq:rhospr} to hold is to choose $\mu$ such that
\begin{equation}
	\label{eq:stepsize1}
	\begin{split}
		0 < \mu < \frac{2}{\max_k\{\lambda_{\max}(\bR_k)\}+2\,\eta}        
	\end{split}
\end{equation}
where $\lambda_\text{max}(\cdot)$ denotes the maximum eigenvalue of its matrix argument. In that case, it follows from \eqref{eq:Ev} that the asymptotic mean bias is given by
\begin{equation}
	\label{eq:bias}
	\lim_{n\rightarrow\infty}E\{\bv(n)\} = \mu\,\eta\,\left\{ \bA_I^\top \left[ \bI_{LN} - \mu\, (\bH_R + \eta \, \bQ) \right]- \bI_{LN} \right\}^{-1}\bA_I^\top\bQ\, \bw^\star.
\end{equation}

\begin{proof}
% As the second term on the right hand side of~\eqref{eq:Ev} is constant, the convergence of this recursion requires the spectral radius of the matrix $\left(\bA_I^\top \left(\bI_{LN} - \mu\, (\bH_R + \eta \, \bQ)\right)\right)$ to be strictly smaller than one, that is,
%\begin{equation}
%	\rho \left(\bA_I^\top \left(\bI_{LN} - \mu\, (\bH_R + \eta \, \bQ)\right)\right)<1.
%\end{equation}
Since any induced matrix norm is lower bounded by the spectral radius, we have the following relation in terms of the block maximum norm (see~\cite{Sayed2013intr} for definition and properties of the norm):
\begin{equation}
	\label{eq:spr}
	\begin{split}
	\rho \left(\bA_I^\top \left[\bI_{LN} - \mu\, (\bH_R + \eta \, \bQ)\right]\right) \leq \|\bA_I^\top \left(\bI_{LN} - \mu\, (\bH_R + \eta \, \bQ)\right)\|_{b,\infty}
	\end{split}
\end{equation}
Now using norm inequalities and the fact that $\bA$ is a left-stochastic matrix (whose block maximum norm is equal to one), we find that:
\begin{equation}
         \label{eq:normineqs}
	\begin{split}
	\|\bA_I^\top \left[\bI_{LN} - \mu\, (\bH_R + \eta \, \bQ)\right]\|_{b,\infty}
	&\leq \|\bA_I^\top\|_{b,\infty} \cdot \| \bI_{LN} - \mu\, (\bH_R + \eta \, \bQ) \|_{b,\infty}  \\
	& = \| \bI_{LN} - \mu\, (\bH_R + \eta \, \bQ)\|_{b,\infty} \\
	& \leq \| \bI_{LN} - \mu\, \bH_R - \mu\eta \bI_{LN}\|_{b,\infty}+ \mu\eta \, \|\bP \otimes \bI_L\|_{b,\infty}
	\end{split}
\end{equation}
using the definition  $\bQ = (\bI_N - \bP) \otimes \bI_L$ and the triangle inequality. Now, it holds that
\begin{equation}
         \label{eq:normP}
	\begin{split}
		\|\bP \otimes \bI_L\|_{b,\infty}= \| \bP\|_{\infty} = 1
	\end{split}
\end{equation}
because $\bP$ is a right stochastic matrix according to condition~\eqref{eq:rho}. Furthermore, since $(1-\mu\eta)\,\bI_{LN} - \mu\,\bH_R$ is a block diagonal Hermitian matrix, its block maximum norm is equal to  its spectral radius~\cite{Sayed2013intr}, namely,
\begin{equation}
          \label{eq:maxrho}
	\|(1-\mu\eta)\,\bI_{LN} - \mu\,\bH_R\|_{b,\infty} = \rho\left((1-\mu\eta)\,\bI_{LN} - \mu\,\bH_R\right).
\end{equation}
Using~\eqref{eq:normP}--\eqref{eq:maxrho} in~\eqref{eq:normineqs} we conclude that a sufficient condition for mean stability is to require
\begin{equation}
	\rho\left((1-\mu\eta)\,\bI_{LN} - \mu\,\bH_R\right) + \mu\eta \leq 1,
\end{equation}
which yields condition~\eqref{eq:stepsize1}.
\end{proof}
%\cblue{As usual, the regularization leads to a biased estimate. However a proper selection of regularizer and its parameter can be expected to decrease significantly error variance at the cost of a small level of bias, so as to reduce the mean-square error. In what follows we study the mean-square error performance for the algorithm.}

\subsection{Mean-square error behavior analysis}

In order to make the presentation clearer, we shall use the following notation for terms in the weight-error expression~\eqref{eq:v}:
\begin{equation}
	\begin{split}
             \bB(n) &=  \bA_I^\top \left[\bI_{LN} - \mu\, (\bH_x(n) + \eta \, \bQ)\right]   \\
             \bg(n) & = \bA_I^\top\,\bp_{zx}(n)  \\
             \br &=  \bA_I^\top\bQ\,\bw^\star
	\end{split}
\end{equation}
so that
\begin{equation}
        \bv(n+1) = \bB(n)\,\bv(n) + \mu\,\bg(n) - \mu\,\eta\,\br.
\end{equation}
Using  Assumption 1 and $E\{\bg(n)\}=0$, the mean-square of the weight error vector $\bv(n+1)$, weighted by any positive semi-definite matrix $\bSig$ that we are free to choose, satisfies the following relation:
\begin{equation}
	\label{eq:EvSig}
	\begin{split}
	E\{\|\bv(n+1)\|^2_{\bSig}\} 	&= E\{\|\bv(n)\|^2_{\bSig'}\} + \mu^2\,\, \tr \left\{ \bSig  E\{\bg(n)\bg^\top(n)\} \right\} + \mu^2\eta^2 \, \|\br\|^2_{\bSig}  \\
 						&\qquad- 2\mu\,\eta\,\br^\top\,\bSig\bB\,E\{\bv(n)\}
	\end{split}
\end{equation}
where
\begin{align}
         &\bB \triangleq\{\bB(n)\} = \bA_I^\top \left[(\bI_{LN} - \mu\, (\bH_R + \eta \, \bQ)\right] \\
         &\bSig' \triangleq E\{ \bB^\top(n)\,\bSig\,\bB(n)\}
 \end{align}
 In expression~\eqref{eq:EvSig}, \cblue{the} freedom in selecting  $\bSig$ will allow us to derive several performance metrics. % via a proper selection, and $\bSig' = E\{ \bB^\top(n)\,\bSig\,\bB(n)\}$. \cblue{It can be seen that the first two terms on the right-hand-side represents the unbiased error, while the last two terms involving $\br$ describe the extra error due to the bias caused by the regularization.\footnote{\cmag{Substracting the bias from \eqref{eq:bias} yields an unbiased error that consists equivalently of excluding the terms depending on $\br$ of \eqref{eq:EvSig}.}} } Let us denote by $\bG$ the following expected value
 Let
\begin{equation}
	\begin{split}
	\bG 	& = E\{\bg(n)\bg^\top(n)\}  \\
		& = \bA_I^\top \,\bC_I^\top\,\text{diag}\{\sigma_{z,1}^2 \bR_{x,1},\dots,\sigma_{z,N}^2 \bR_{x,N}\}\,\bC_I\,\bA_I
	\end{split}
\end{equation}
where $\bC_I = \bC\otimes \bI_L$. Then, relation~\eqref{eq:EvSig} can be rewritten as
\begin{equation}
	\label{eq:EvSig2}
	E\{\|\bv(n+1)\|^2_{\bSig}\} 	= E\{\|\bv(n)\|^2_{\bSig'}\} + \mu^2\,\, \tr \left\{\bSig\bG\right\} + \mu^2\eta^2 \, \|\br\|^2_{\bSig}
 						- 2\mu\,\eta\,\br^\top\,\bSig\bB\,E\{\bv(n)\}
\end{equation}
We would like to show that this variance relation converges for sufficiently small step-sizes and we would also like to evaluate its steady-state value in order to determine the mean-square-error of the distributed strategy. However, note that the weighting matrices $\bSig$ and $\bSig'$ on both sides of~\eqref{eq:EvSig2} are different, which means that~\eqref{eq:EvSig2} is still not an actual recursion. To handle this situation, we transform the weighting matrices into vector forms as follows. 
Let $\vc(\cdot)$ denote the operator that stacks the columns of a matrix on top of each other. Vectorizing the matrices $\bSig$ and $\bSig'$ by $\bsig=\vc(\bSig)$ and $\bsig' = \vc(\bSig')$, it can be verified that the relation between them can be expressed as the following linear transformation:
\begin{equation}
	\bsig' = \bK \, \bsig
\end{equation}
where $\bK$ is the $(LN)^2 \times (LN)^2$ matrix given by
\begin{equation}
	\begin{split}
	\bK	& = E\{\bB^\top(n)\otimes\bB^\top(n)\} \\
		& = \bA_I \otimes \bA_I - \mu\,(\bH_R+\eta\,\bQ)^\top \bA_I \otimes \bA_I - \mu\,\bA_I \otimes  (\bH_R+\eta\,\bQ)^\top\bA_I  \\
		& \qquad+ \mu^2 \, E\{ (\bH(n)+\eta\,\bQ)^\top \bA_I\otimes  (\bH(n)+\eta\,\bQ)^\top \bA_I \}.
	\end{split}
\end{equation}
Neglecting the influence of second-order terms in $\mu$, $\bK$ can be approximated by
\begin{equation}
	\label{eq:Kapp}
	\bK \approx \bB^\top \otimes \bB^\top.
\end{equation}
Finally, let us define $\f(\bsig, E\{\bv(n)\})$ as the last two terms on the right hand side of~\eqref{eq:EvSig2}, i.e.,
\begin{equation}
	\label{eq:deff}
	\f(\bsig, E\{\bv(n)\}) \triangleq  \mu^2\eta^2 \, \|\br\|^2_{\bsig} - 2\mu\,\eta\, \left(\bB\,E\{\bv(n)\} \otimes \br\right)^\top\, \bsig.
\end{equation}
For notational convenience, we \cblue{are omitting} the argument $\br$ of $\f$ \cblue{since} it is deterministic. Equation~\eqref{eq:EvSig2} can be expressed as follows:
\begin{equation}
	\label{eq:EvSig3}
	\begin{split}
		E\{\|\bv(n+1)\|^2_{\bsig}\} &= E\{\|\bv(n)\|^2_{\bK\bsig}\} + \mu^2\,\, \vc(\bG^\top)^\top\,\bsig + \f(\bsig, E\{\bv(n)\})
	\end{split}
\end{equation}
where we will be using the notations $\|\!\cdot\!\|_\Sigma$ and $\|\!\cdot\!\|_\sigma$ interchangeably.

\theorem (Mean-square stability) Assume data model~\eqref{eq:datamodel} and Assumption 1 hold. Assume further that the step-size $\mu$ is sufficiently small such that approximation~\eqref{eq:stepsize1} is justified by neglecting higher-order powers of $\mu$, and relation~\eqref{eq:EvSig3} can be used as a reasonable representation for the evolution of the (weighted) mean-square-error. Then the diffusion multitask strategy~\eqref{eq:ATC_MSEl2} is mean-square stable if the matrix $\bK$ is stable. Under approximation~\eqref{eq:Kapp}, the stability of $\bK$ is guaranteed by sufficiently small step-sizes that also satisfy~\eqref{eq:stepsize1}.

\begin{proof}
Iterating recursion~\eqref{eq:EvSig3} starting from $n=0$, we find that
\begin{equation}
	\label{eq:EVn0}
	E\{\|\bv(n+1)\|^2_{\bsig}\} = E\{\|\bv(0)\|^2_{\bK^{n+1}\bsig}\} + \mu^2\,\, \vc(\bG^\top)^\top\,\sum_{i=0}^n
	\bK^i \bsig + \sum_{i=0}^n\f(\bK^i\bsig, E\{\bv(n-i)\})
\end{equation}
with initial condition $\bv(0)=\bw(0)-\bw^\star$. \cblue{Provided that $\bK$ is stable, the first and the second term on the RHS of \eqref{eq:EVn0} converge as $n\rightarrow \infty$, to zero for the former, and to a finite value for the latter. Consider now the third term on the RHS of \eqref{eq:EVn0}. We know from~\eqref{eq:Ev} that $E\{\bv(n)\}$ is uniformly bounded because \eqref{eq:Ev} is a BIBO stable recursion with a bounded driving term $- \mu\,\eta\,\bA_I^\top\bQ\, \bw^\star$. Moreover, from~\eqref{eq:deff}, the expression for $\f(\bK^i\bsig, E\{\bv(n-i)\})$ can be written as
\begin{equation}
       \f(\bK^i\bsig, E\{\bv(n-i)\}) =  \mu^2\eta^2 \, \vc\{\br\br^\top\}^\top \bK^i \bsig - 2\mu\,\eta\left(\bB\,E\{\bv(n-i)\} \otimes \br\right)^\top\bK^i\bsig.
\end{equation}
We further know that $\bK$ is stable. Therefore, there exists a matrix norm~\cite{Sayed2013intr}, denoted by $\|\!\cdot\!\|_{\rho}$ such that $\|K\|_{\rho}=c_\rho<1$. Applying this norm to $\f$ and using the triangular inequality, we can deduce that 
\begin{equation}
      | \f(\bK^i\bsig, E\{\bv(n-i)\}) | < \nu\,c_\rho^i
\end{equation}
for some positive finite constant $\nu$. It follows that the sum appearing as the right-most term in (68) converges as $n\rightarrow\infty$. We conclude that $E\{\|\bv(n+1)\|^2_{\sigma}\}$ converges to a bounded value as $n\rightarrow\infty$, and the algorithm is said to be mean-square stable.}
\end{proof}

\theorem (Transient MSD) Considering a sufficiently small step-size $\mu$ that ensures mean and mean-square stability, and selecting $\bSig = \frac{1}{N}\bI_{LN}$, then the network MSD learning curve, defined by $\zeta(n)=\frac{1}{N}E\{\|\bv(n)\|\}^2$ evolves according to  the following recursions for~$n\geq 0$:
\begin{align}
	\begin{split}\label{eq:TransMSD1}
		\zeta(n+1)  	&= \zeta(n) + \frac{1}{N} \Big( \mu^2\, \vc(\bG^\top)^\top\, \bK^{n} \vc(\bI_{LN})
			- E\{\|\bv(0)\|^2_{(\bI_{(NL)^2}-\bK)\bK^n\vc(\bI_{LN})}\} +{\mu^2\eta^2} \|\br\|^2_{\bK^n\vc(\bI_{LN})}  \\
					&\hspace{0.5cm} -2\mu\eta\,(\bGam(n) + \left( \bB\,E\{\bv(n)\} \otimes \br \right)^\top \vc(\bI_{LN})\Big)   \end{split} \\
		\bGam(n+1) &= \bGam(n)\, \bK +   \left(\bB\,E\{\bv(n)\}\otimes \br\right)^\top (\bK-\bI_{(LN)^2})   \label{eq:TransMSD2}
\end{align}
with initial condition $\zeta(0) = \frac{1}{N}\|\bv(0)\|^2$ and $\bGam(0)  = \cb{0}_{(LN)^2}$.

\begin{proof}
Comparing~\eqref{eq:EVn0} at instants $n+1$ and $n$, we can relate $E\{\|\bv(n+1)\|^2_{\bsig}\}$ to $E\{\|\bv(n)\|^2_{\bsig}\} $ as follows:
\begin{equation}
	\label{eq:EVrec_sig}
		\begin{split}
			E\{\|\bv(n+1)\|^2_{\bsig}\} = &E\{\|\bv(n)\|^2_{\bsig}\} + \mu^2\,\, \vc(\bG^\top)^\top\, \bK^{n} \bsig
			- E\{\|\bv(0)\|^2_{(\bI_{(NL)^2}-\bK)\bK^n\bsig}\}\\
                 &+\sum_{i=0}^n\f(\bK^i\bsig, E\{\bv(n-i)\}) -\sum_{i=0}^{n-1}\f(\bK^i\bsig, E\{\bv(n-1-i)\})
		\end{split}
\end{equation}
\cblue{We can} rewrite the last two terms on the RHS of~\eqref{eq:EVrec_sig} as follows:
\begin{equation}
	\begin{split}
		\sum_{i=0}^n\f(\bK^i\bsig&, E\{\bv(n-i)\}) - \sum_{i=0}^{n-1}\f(\bK^i\bsig, E\{\bv(n-1-i)\})  \\
		& = \mu^2\eta^2 \, \|\br\|^2_{\bK^n\bsig} - \sum_{i=0}^n \left\{  2\mu\,\eta\,  \left(\bB\,E\{\bv(n)\}\otimes \br\right)^\top\, \bK^i\bsig \right\} \\
		& \qquad + \sum_{i=0}^{n-1} \left\{ 2\mu\,\eta\,  \left(\bB\,E\{\bv(n)\}\otimes \br\right)^\top\, \bK^i\bsig \right\} \\
		& = \mu^2\eta^2 \, \|\br\|^2_{\bK^n\bsig}  -2\mu\,\eta\,  \left(\bB\,E\{\bv(n)\}\otimes \br\right)^\top \bsig \\
		& \qquad-2\mu\,\eta\,  \Big\{  \sum_{i=1}^n  \left(\bB\,E\{\bv(n)\}\otimes \br\right)^\top\, \bK^i  + \sum_{i=0}^{n-1}  
		 \left(\bB\,E\{\bv(n)\}\otimes \br\right)^\top\, \bK^i \Big\} \bsig.
         \end{split}
\end{equation}
Introducing the following notation
\begin{equation}
        \bGam(n) = \sum_{i=1}^n \left(\bB\,E\{\bv(n)\}\otimes \br\right)^\top\, \bK^i  + \sum_{i=0}^{n-1}   \left(\bB\,E\{\bv(n)\}\otimes \br\right)^\top\, \bK^i,
\end{equation}
we can reformulate recursion~\eqref{eq:EVrec_sig} as follows:
\begin{align}
                     \begin{split}
                               E\{\|\bv(n+1)\|^2_{\bsig}\} &= E\{\|\bv(n)\|^2_{\bsig}\} + \mu^2\,\, \vc(\bG^\top)^\top\, \bK^{n} \bsig
                               - E\{\|\bv(0)\|^2_{(\bI_{(NL)^2}-\bK)\bK^n\bsig}\} + \mu^2\eta^2\|\br\|^2_{\bK^n\bsig}  \\
                                &\hspace{0.5cm} -2\mu\eta\,(\bGam(n) +  \left(\bB\,E\{\bv(n)\}\otimes \br\right)^\top \bsig   \end{split} \label{eq:MSDsigma1} \\
                \bGam(n+1) &= \bGam(n) \bK +   \left(\bB\,E\{\bv(n)\}\otimes \br\right)^\top (\bK-\bI_{(LN)^2})  \label{eq:MSDsigma2}
\end{align}
with $\bGam(0)=\cb{0}_{(LN)^2}$. To derive the transient curve for the MSD, we replace $\sigma$ by $\frac{1}{N}\vc(\bI_{LN})$.

\end{proof}

\theorem (Steady-state MSD) If the step size is chosen sufficiently small to ensure mean and mean-square-error convergence, then the value of the steady-state MSD  for the diffusion network~\eqref{eq:ATC_MSEl2} is given by
\begin{equation}
	\label{eq:MSD}
	\zeta^\star=  \frac{\mu^2}{N}\, \vc(\bG^\top)^\top\, (\bI_{(LN)^2} - \bK)^{-1} \vc(\bI_{LN})
		+ \f\left(\frac{1}{N}(\bI_{(LN)^2} - \bK)^{-1} \vc(\bI_{LN}), E\{\bv(\infty)\}\right)
\end{equation}
where $E\{\bv(\infty)\}$ is determined by expression~\eqref{eq:bias}.

\begin{proof} The steady-state MSD is the limiting value
\begin{equation}
	\label{eq:MSD.def}
	\zeta^\star = \lim_{n\rightarrow\infty}\frac{1}{N}E\{\|\bv(n)\|\}^2.
\end{equation}
From the recursive expression~\eqref{eq:EvSig3} we obtain as  $n\rightarrow\infty$ that
\begin{equation}
	\label{eq:stablepoint}
	\lim_{n\rightarrow\infty}E\{\|\bv(n)\|^2_{(\bI_{(NL)^2}-\bK)\bsig}\} =  \mu^2\,\, \vc(\bG^\top)^\top\,\bsig + \f(\bsig, E\{\bv(\infty)\}).
\end{equation}
Comparing expressions~\eqref{eq:MSD.def} and~\eqref{eq:stablepoint}, we observe that to arrive at the MSD requires us to choose $\sigma$ to satisfy
\begin{equation}
	\label{eq:sig_MSD}
	(\bI_{(NL)^2} - \bK) \,\bsig_{\text {MSD}} = \frac{1}{N} \vc(\bI_{LN}).
\end{equation}
This leads  to expression~\eqref{eq:MSD}.
\end{proof}

 %=================  Simulation ==========================

\section{Simulation examples}

In this section, we first conduct simulations on a simple network to illustrate the proposed algorithm and the analytical performance models. Then, we  provide several application-oriented examples where the proposed distributed learning strategy may find applications. These experiments will illustrate the behavior and the advantage of the proposed strategy.

\begin{figure*}[!t]
	\subfigure{
   	\begin{minipage}[c]{.5\linewidth}
   		\centering
      		\includegraphics[scale=0.45]{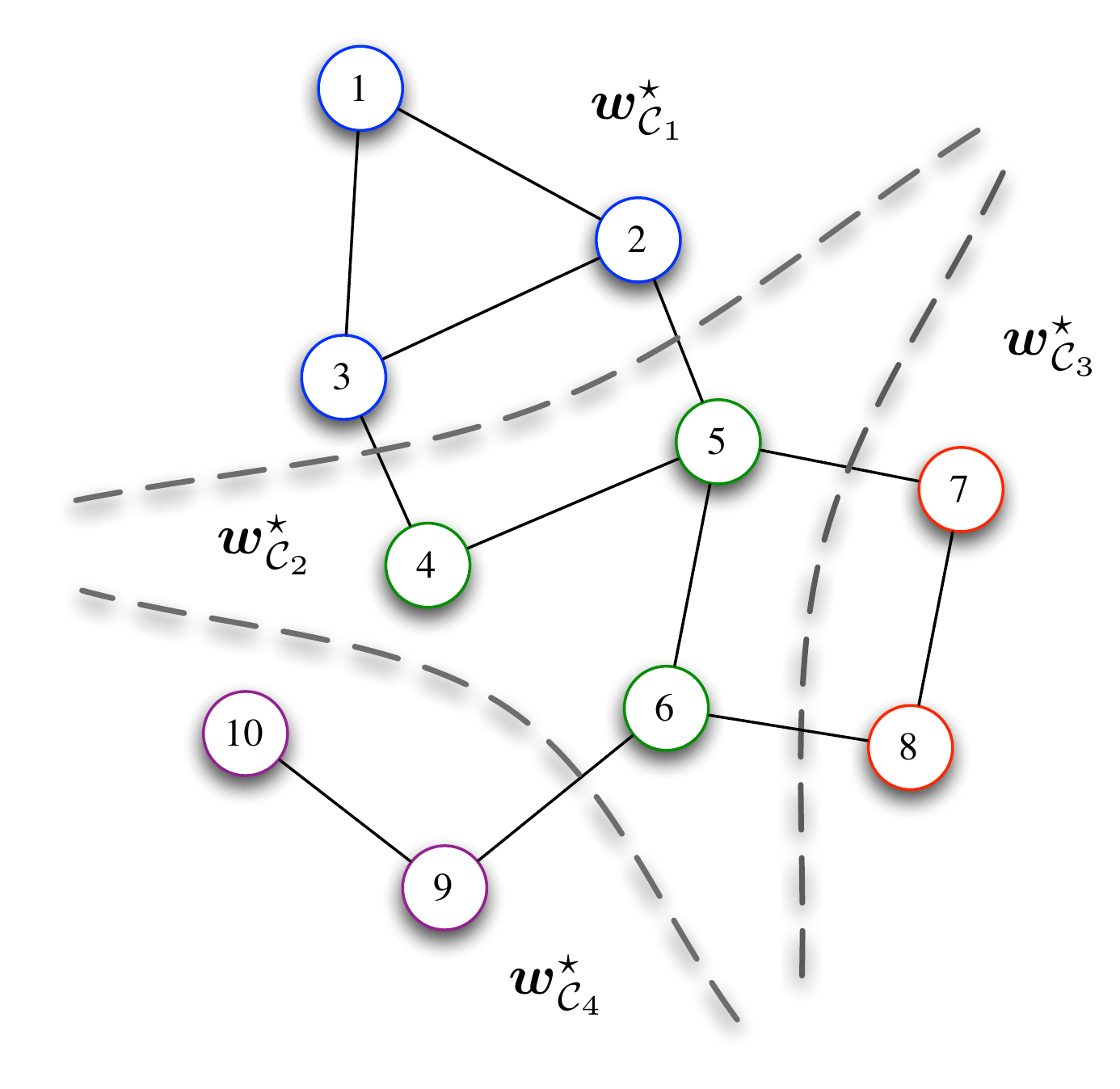}
   	\end{minipage}} \hfill
	\subfigure{
   		\begin{minipage}[c]{.5\linewidth}
   		\centering
      		\includegraphics[scale=0.45]{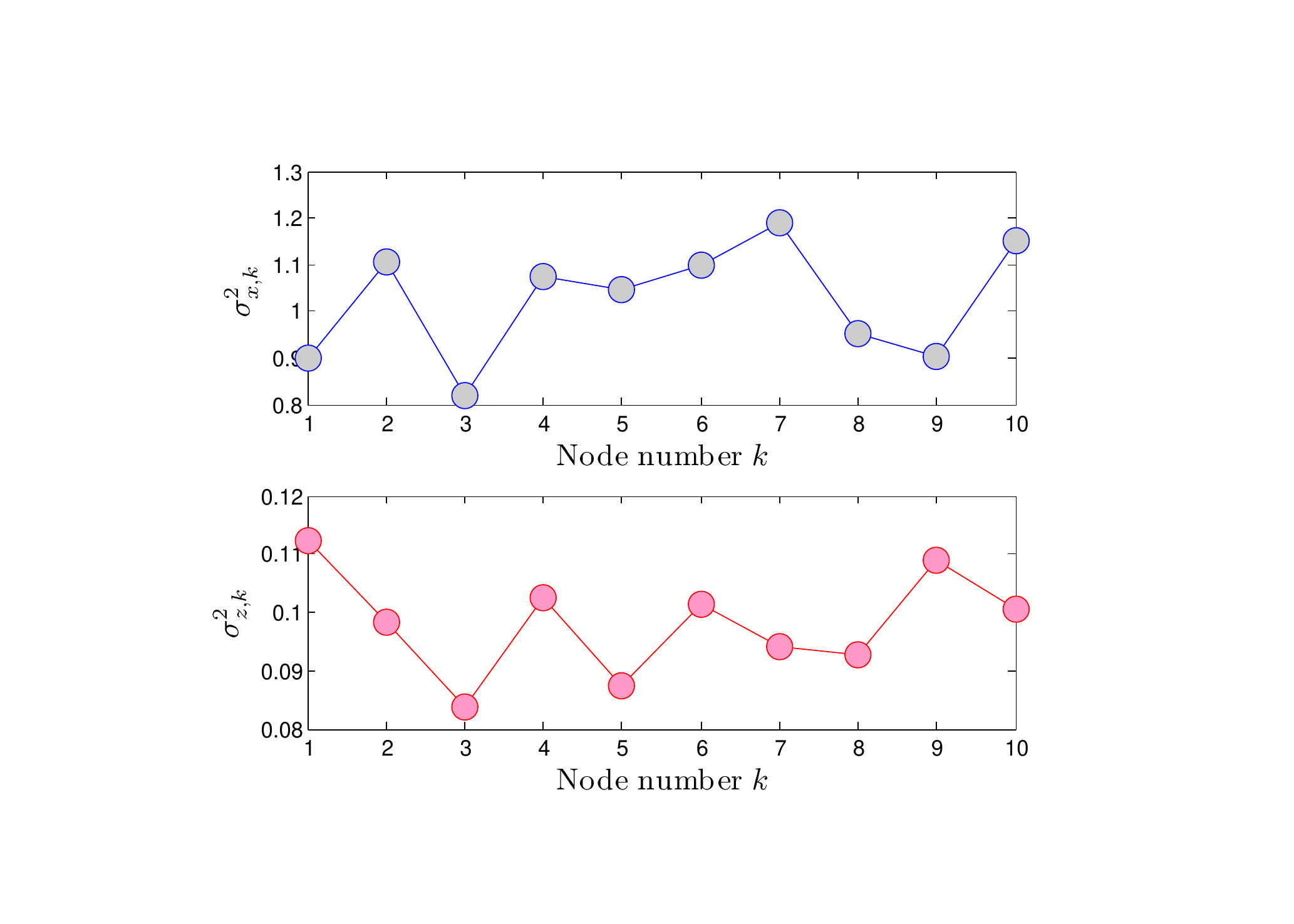}
   	\end{minipage}}
	\caption{Experimental setup. Left: network studied in Section~\ref{sec:simu1}, with $10$ nodes divided into 4 different clusters. Right: input
	signal and noise variances for each node.}
	\label{fig:simu1}
\end{figure*}

\subsection{Illustrative numerical example}
\label{sec:simu1}

In this subsection we provide an illustrative  example to show how the proposed distributed algorithm converges over clustered multitask network. We consider a network consisting of $10$ nodes with the topology depicted in Figure~\ref{fig:simu1}~(left). The nodes were divided into $4$ clusters:
$\C_1 = \{1, 2, 3\}$, $\C_2 = \{4, 5, 6\}$, $\C_3 = \{7, 8\}$ and $\C_4 = \{9, 10\}$. Two-dimensional coefficient vectors of the form $\bw^\star_{\C_i}=\bw_o+\delta\bw_{\C_i}$ were chosen as $\bw_o = [0.5, - 0.4]^\top$, $\delta\bw_{\C_1} = [0.0287,-0.005]^\top$, $\delta\bw_{\C_2} = [0.0234,0.005]^\top$, $\delta\bw_{\C_3} =[-0.0335, 0.0029]^\top$, and $\delta\bw_{\C_4} = [0.0224, 0.00347]^\top$. The regression inputs $\bx_{k}(n)$ were zero-mean $2\times 1$ random vectors governed by a Gaussian distribution with covariance matrices $\bR_{x,k} = \sigma_{x,k}^2\,\bI_L$, and the $\sigma_{x,k}^2$ shown in the top right plot of Figure~\ref{fig:simu1}. The background noises $z_k(n)$ were i.i.d. zero-mean Gaussian random variables, independent of any other signals. The corresponding variances $\sigma_{z,k}^2$ are depicted in the bottom right plot of Figure~\ref{fig:simu1}.

Regularization strength $\rho_{k\ell}$ was chosen as $\rho_{k\ell} = |\N{k}\backslash\C(k)|^{-1}$ for $\ell\in\N{k}\backslash \C(k)$, and  $\rho_{k\ell} = 0$ for any other $\ell$. This setting usually leads to asymmetrical regularization weights. We considered the diffusion algorithm with measurement diffusion governed by a uniform matrix $\bC$ such that $c_{\ell k} = |\N{\ell}\cap\C(\ell)|^{-1}$ for $k \in \N{\ell}\cap\C(\ell)$. Likewise, a uniform $\bA$ was used such that  $a_{\ell k} = |\N{k}\cap\C(k)|^{-1}$ for $\ell \in \N{k}\cap\C(k)$.

\begin{figure*}[!h]
	\subfigure{
   	\begin{minipage}[c]{.5\linewidth}
   		\centering
      		\includegraphics[trim = 18mm 0mm 25mm 10mm, clip, scale=0.5]{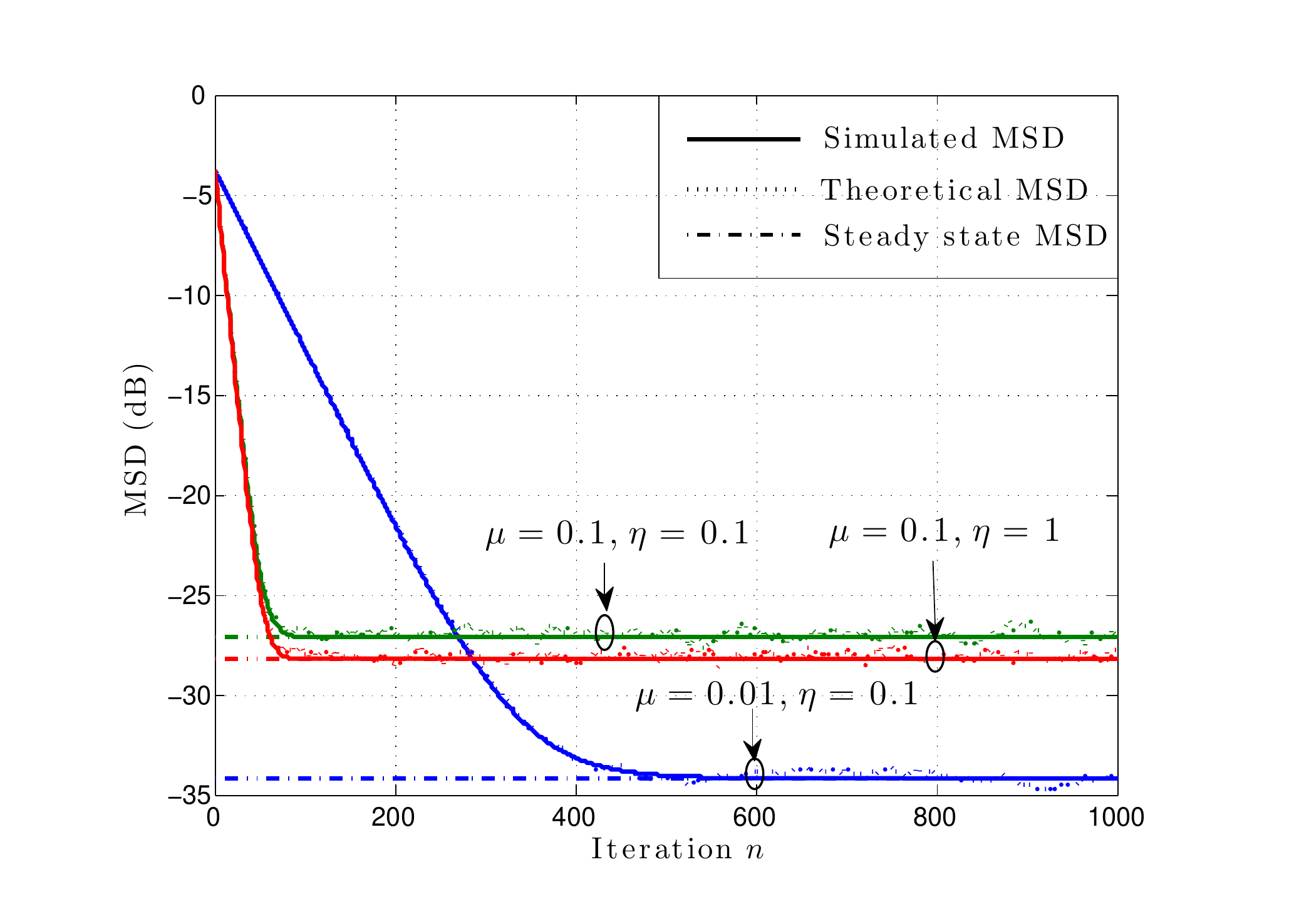}
   	\end{minipage}} \hfill
	\subfigure{
   		\begin{minipage}[c]{.5\linewidth}
   		\centering
      		\includegraphics[trim =18mm 0mm 25mm 10mm, clip, scale=0.5]{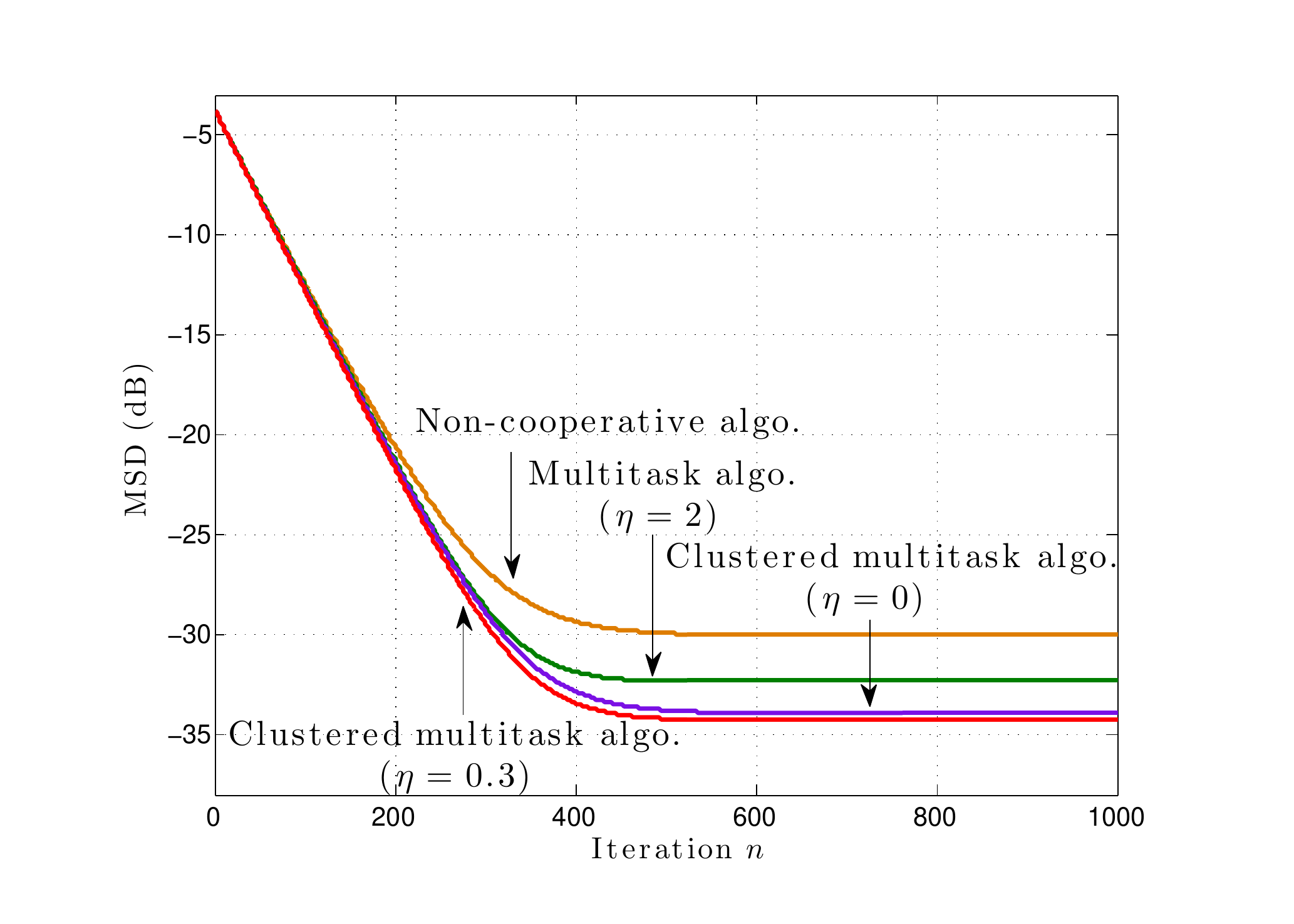}
   	\end{minipage}}
% 	\subfigure{
%   	\begin{minipage}[c]{.315\linewidth}
%   		\centering
%      		\includegraphics[trim = 15mm 75mm 15mm 85mm, clip, scale=0.37]{MSD_cluster.pdf}
%   	\end{minipage}} \hfill
	\caption{Network performance illustration. Left: transient and steady-state MSD (model vs. Monte Carlo) for different step-sizes and regularization
	parameters. Right: performance comparison for different strategies using theoretical models.}% Right: MSD evolution in each cluster using theoretical models.}
	\label{fig:simu1res}
\end{figure*}

The algorithm was run with different step-size and regularization parameters $(\mu,\eta)$ such as $(0.01, 0.1)$, $(0.05, 0.1)$ and $(0.01, 1)$. Simulation results were obtained by averaging $100$ Monte-Carlo runs. Transient MSD curves were obtained by~\eqref{eq:TransMSD1} and~\eqref{eq:TransMSD2}. Steady-state MSD values were obtained by  expression~\eqref{eq:MSD}. It can be observed in the left plot of Figure~\ref{fig:simu1res} that the models accurately match the simulated results.

These models were used to illustrate the performance of several learning strategies: 1) the non-cooperative LMS algorithm, 2) the multitask algorithm and 3) the clustered multitask algorithm. The non-cooperative algorithm was obtained by assigning a cluster to each node and setting $\eta=0$. The multitask algorithm was obtained by assigning a cluster to each node and setting $\eta\neq 0$. The right plot of Figure~\ref{fig:simu1res} shows that the noncooperative algorithm has the largest MSD as nodes do not collaborate for additional benefit.  If estimation is performed without cluster information, but only with regularization between nodes as in the case of the multitask diffusion LMS, it can be observed that the performance is better than in the non-cooperative case. Finally, providing prior information to the clustered multitask network via an appropriate definition of clusters leads to the best performance. Clustering strategies are not discussed in this paper. This will be investigated in future work. One strategy is proposed in \cite{Zhao2012}.

\subsection{Distributed spectrum estimation with multi-antenna devices}

We now consider an example of distributed spectral sensing using the clustered multitask diffusion LMS. Cognitive radio systems involve two types of users: primary users (PU) and secondary users (SU). Secondary users are allowed to detect and temporarily occupy unused frequency bands provided that they do not cause harmful interference to primary users~\cite{Mitola1999cognitive}. Therefore, secondary users need to sense spectral bands that are occupied by active primary users. In~\cite{Sayed2013intr,Lorenzo2013Cog},  collaborative spectral sensing was studied  with single-antenna nodes via the diffusion strategy. In this subsection,  we explore a distributed spectral sensing method over the network with multi-antenna devices.

Consider a communication environment consisting of $N_P$ primary users and $N_S$ secondary users, each secondary user being equipped with $N_R$ antennas. We assume that the power spectrum of the signal transmitted by each primary user $q$ can be represented as a linear combination of $N_B$ basis functions weighted by the weights $\alpha_{qm}$, namely,
\begin{equation}
	\cb{S}_q(f) =   \sum_{m=1}^{N_B} \alpha_{qm} \, \cb{\phi}_m(f)
\end{equation}
The power spectrum of the signal received at the $\ell$-th antenna of  device $k$ is given by
\begin{equation}
	\cb{R}_{k\ell}(f) = \sum_{q=1}^{N_P}  p_{k\ell,q} \,\left(\sum_{m=1}^{N_B} \alpha_{qm} \, \cb{\phi}_m(f)\right) + \sigma_{k\ell}^2
\end{equation}
where $p_{k\ell,q}$ is the path loss factor between the primary user $q$ and the $\ell$-th antenna of node $k$, and $\sigma_{k\ell}^2$ is the receiver noise power.

At each time instant, the $\ell$-th antenna of device $k$ gets measurements of the power spectrum $\cb{R}_{k\ell}(f)$ over $N_F$ frequency samples. Assume that the receiver noise power can be estimated with high accuracy using energy detection over an idle band. Then, at instant $n$, we can define the reference signal on this antenna element at the $j$-th frequency band by
\begin{equation}
	\label{eq:model_p}
	r_{k\ell,j}(n) = \sum_{q=1}^{N_P}  p_{k\ell,q} \, \left( \sum_{m=1}^{N_B} \alpha_{qm} \, \phi_m(f_j)\right)  
	+ z_{k\ell,j}(n) \quad\text{for}\quad  j = 1, \dots, N_F
\end{equation}
where $z_{k\ell,j}(n)$ is the sampling noise, assumed to be zero-mean Gaussian with variance $\sigma_{z_{k\ell,j}}^2$.  We denote by $\cb{\Phi}$ the $N_F \times N_B$ matrix of basis functions with $j$-th row defined by $[\phi_1(f_j), \ldots, \phi_{N_B}(f_j)]$, and let
 \begin{equation}
	\cb{\Phi}_{k\ell} = [p_{k\ell,1},\dots,p_{k\ell,N_P}] \otimes \cb{\Phi}.
\end{equation}
We collect coefficients $\alpha_{qm}$ into the vectors $\balpha_q=[\alpha_{q1},\ldots,\alpha_{qN_B}]^\top$ and $\balpha=[\balpha^\top_1,\ldots,\balpha^\top_{N_P}]^\top$, measurements $r_{k\ell,j}(n)$ into the vector $\br_{k\ell}(n) = [r_{k\ell,1}(n), \dots, r_{k\ell,N_F}(n)]^\top$, and noise samples $z_{k\ell,j}(n)$ into the vector $\cb{z}_{k\ell}(n) = [z_{k\ell,1}(n), \dots, z_{k\ell,N_F}(n)]^\top$. At time instant $n$, the model~\eqref{eq:model_p} can then be expressed in the following vector form
 \begin{equation}
	\label{eq:model_p_mat}
	\cb{r}_{k\ell}(n) = \cb{\Phi}_{k\ell}\,\balpha + \cb{z}_{k\ell}(n)
\end{equation}
for each antenna $\ell$ of each sensor node $k$, at each time instant $n$. Inverting the linear model \eqref{eq:model_p_mat} should allow each pair $(k,\ell)$ to estimate the solution $\balpha(n)$. However, in practice, the path loss factor $p_{k\ell, q}$ cannot be estimated accurately, or even estimated due to failures of synchronization if the power of the received signal is lower than a certain threshold. Thus, $\hat{\cb{\Phi}}_{k\ell}$, depending on the estimated path loss factors $\hat{p}_{k\ell,q}$, should be used instead of $\cb{\Phi}_{k\ell}$ in the model~\eqref{eq:model_p_mat}. In the experiment described hereafter, we treat each multi-antenna device as a cluster of $N_R$ antennas because they are supposed to sense the same local spectrum. In addition, we assume that each cluster is fully connected, i.e., with links between each pair of antennas. Existence of connections between devices depends on their distance from one another. As a consequence, the number of intra-cluster neighbors for each antenna element is $N_R$, including itself, and the number of extra-cluster neighbors for each antenna element is $N_R$ times the number of neighboring devices. 

\begin{figure}[t]
	\centering
	\includegraphics[scale=0.5]{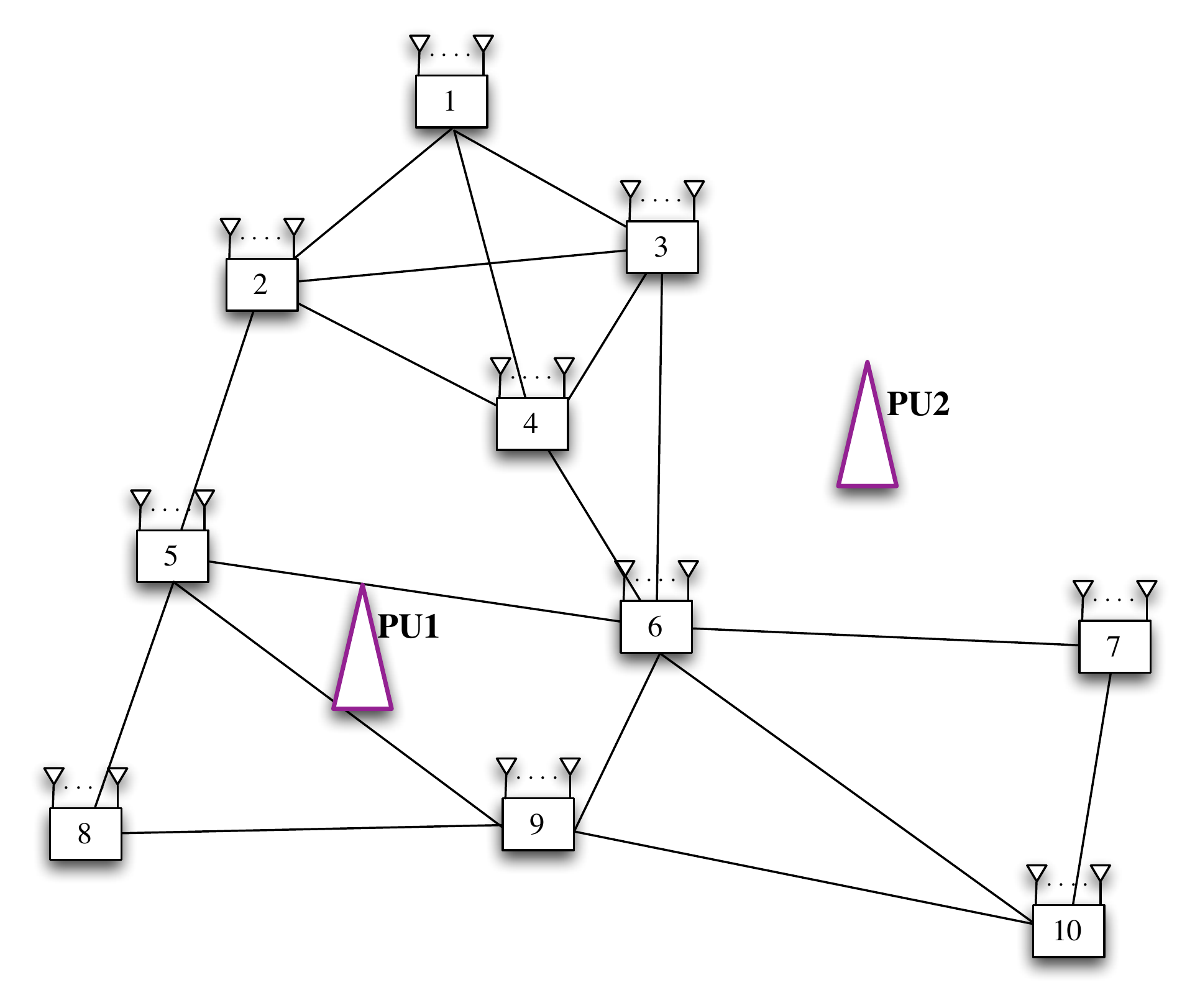}
	\caption{Cognitive radio network with multi-antenna devices.}
	\label{Cognitive_radio}
\end{figure}

Estimating the spectrum in a noncooperative manner would lead to a local profile of the spectrum occupation, with hidden node effects. We used our algorithm to circumvent this drawback. We considered a connected network composed of $N_P= 2$ primary users and $N_S=10$ secondary users. Each secondary user was equipped with an antenna array of $N_R$ elements. The topology of the network is shown in Figure~\ref{Cognitive_radio}. The secondary users sampled $N_F=80$ frequency bins. We used $N_B = 16$ Gaussian basis functions \cblue{defined as
\begin{equation}
       \phi_m(f) = e^{-\frac{\|f-f_{c_i}\|^2}{2\sigma_b^2}}
\end{equation}
with centers} $f_{c_i}$ uniformly distributed along the frequency axis, and variance $\sigma_b^2=0.0025$. Vectors $\balpha_1$ and $\balpha_2$ were arbitrarily set to $\balpha_1 =[\cb{0}^\top_{10} \; 0.4 \; 0.38 \;0.4 \;\cb{0}^\top_3]^\top$ and $\balpha_2 = [\cb{0}^\top_3 \; 0.4 \; 0.38 \; 0.4 \; \cb{0}^\top_{10}]^\top$, where $\cb{0}_q$ is a $q \times 1$ vector of zeros.  The path loss factor at instant $n$ between the primary user $q$ and the $\ell$-th antenna of the secondary user $k$ was set to $p_{k\ell,q}(n) = \bar{p}_{k,q} + \delta p_{k\ell,q}(n) $, where $\bar{p}_{k,q} $ is the deterministic path loss factor determined by the distance between $k$ and $q$ via the free space propagation model, \cblue{i.e., the received signal power is inversely proportional to the squared distance to the transmitter}, and $\delta p_{k\ell,q}(n)$ is a zero-mean Gaussian variable with standard deviation $0.2\,\bar{p}_{k,q}$, which served as the random fading among antenna elements.

In practice, the path loss factors have to be estimated. We considered that each antenna of the secondary user~$k$ was able to estimate $\hat{p}_{k\ell,q}(n) = \bar{p}_{k,q} $ if $\bar{p}_{k,q}\geq p_0$, otherwise $\hat{p}_{k\ell,q}(n)=0$ due to the loss of the synchronization. The noise $z_{k\ell,j}(n)$ was assumed to be zero-mean Gaussian with the standard deviation $0.01$. The antenna elements were considered as fully connected on each device. \cblue{The information exchange matrix $\bC$ and the combination matrix $\bA$ were arbitrarily set to uniform matrices, with entries thus equal to $\frac{1}{N_R}$ as each multi-antenna device was considered as a cluster. Within each cluster, the regularization parameter $\rho_{k\ell}$ was uniformly set to one over $N_R$ multiplied by the number of neighboring devices.}

\begin{figure}[!h]
  		\centering
      		\includegraphics[trim = 20mm 15mm 10mm 0mm, clip,scale=0.48]{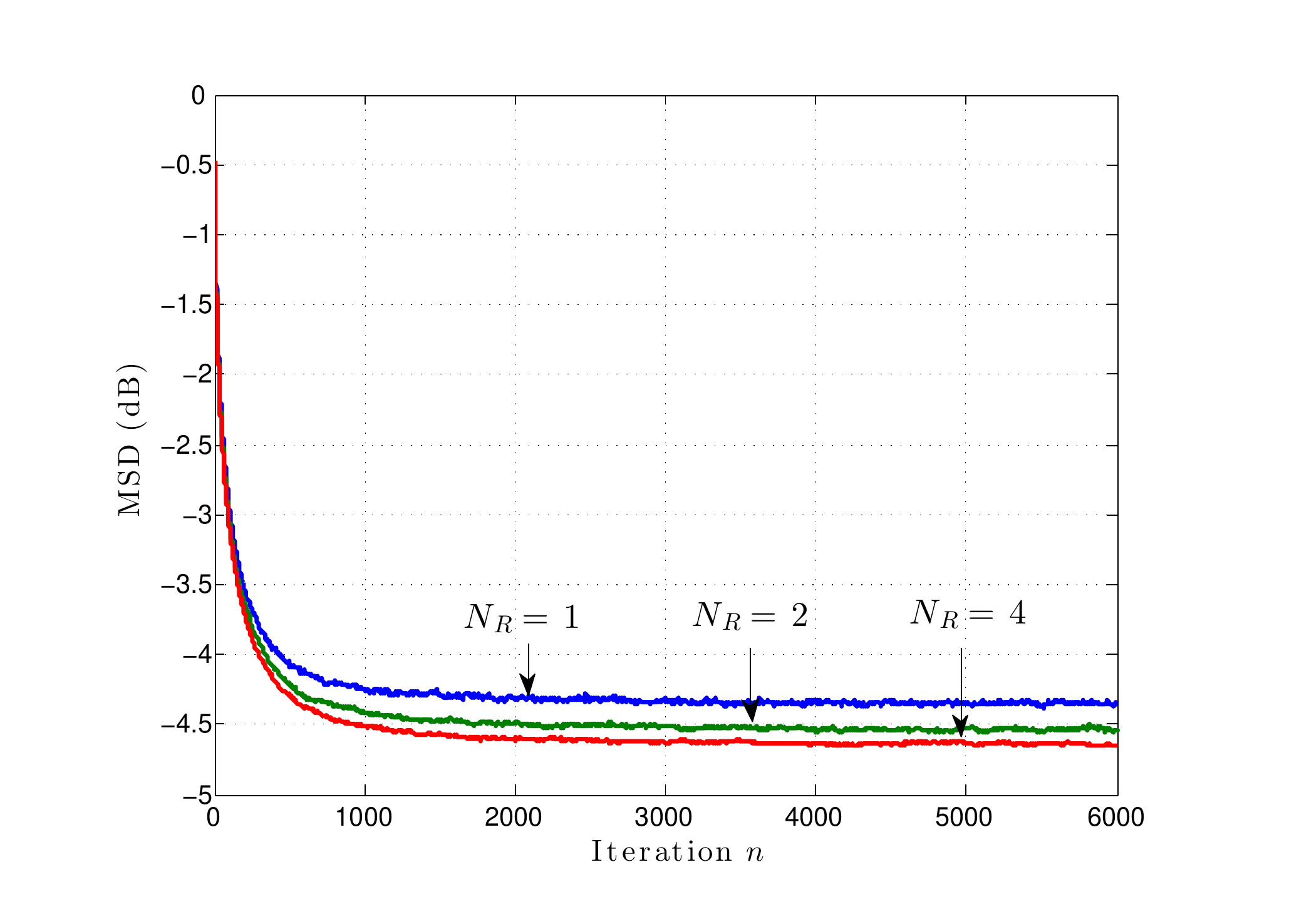}
		\includegraphics[trim = 20mm 15mm 25mm 0mm, clip,scale=0.48]{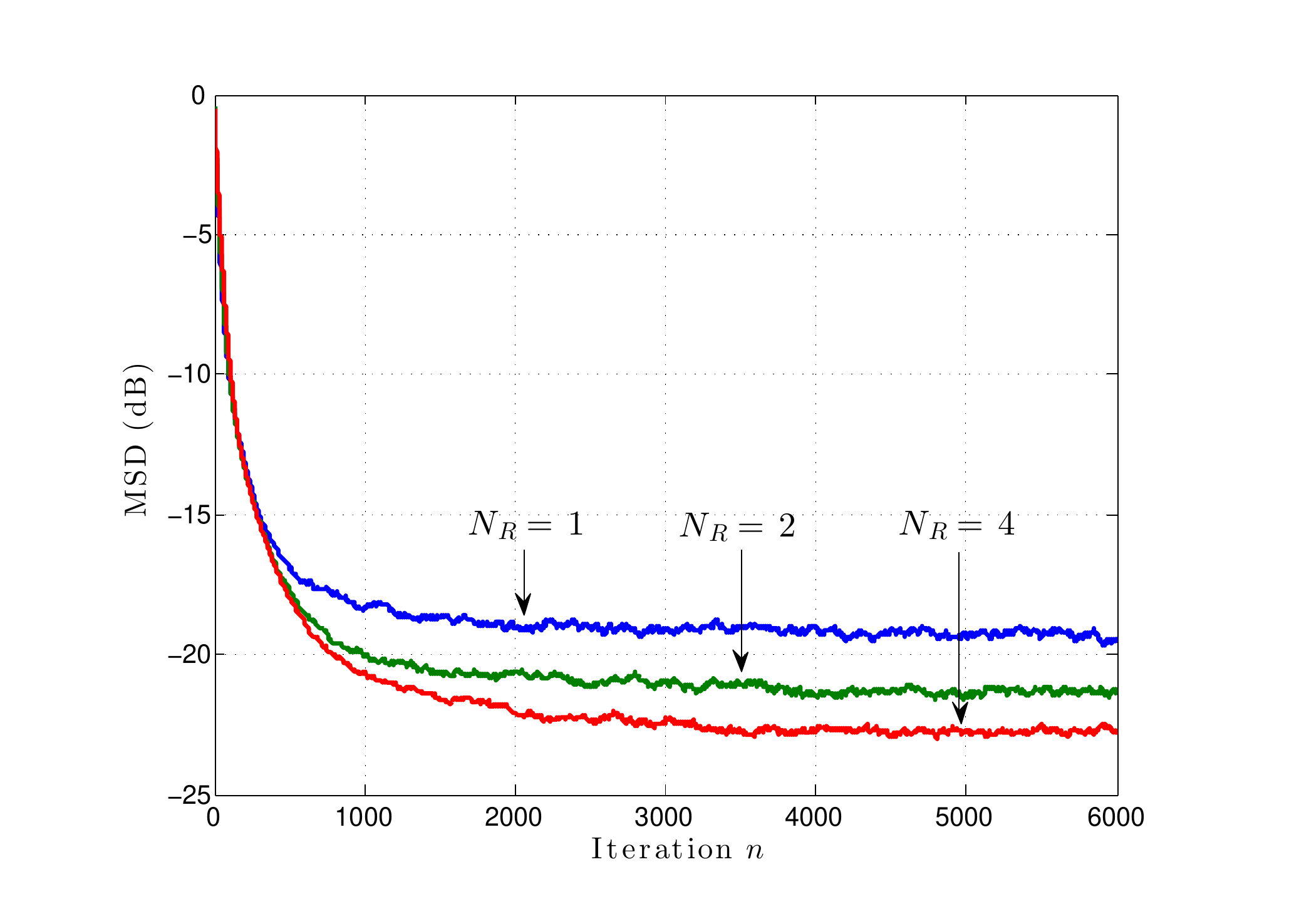}
	\caption{  \label{fig:cog_curve} MSD learning curves. Left: non-cooperating devices $(\eta = 0)$. Right: cooperating devices $(\eta = 0.01)$.}
\end{figure}

\begin{figure}[!h]
  		\centering
      		\includegraphics[trim = 15mm 25mm 15mm 0mm, clip,scale=0.48]{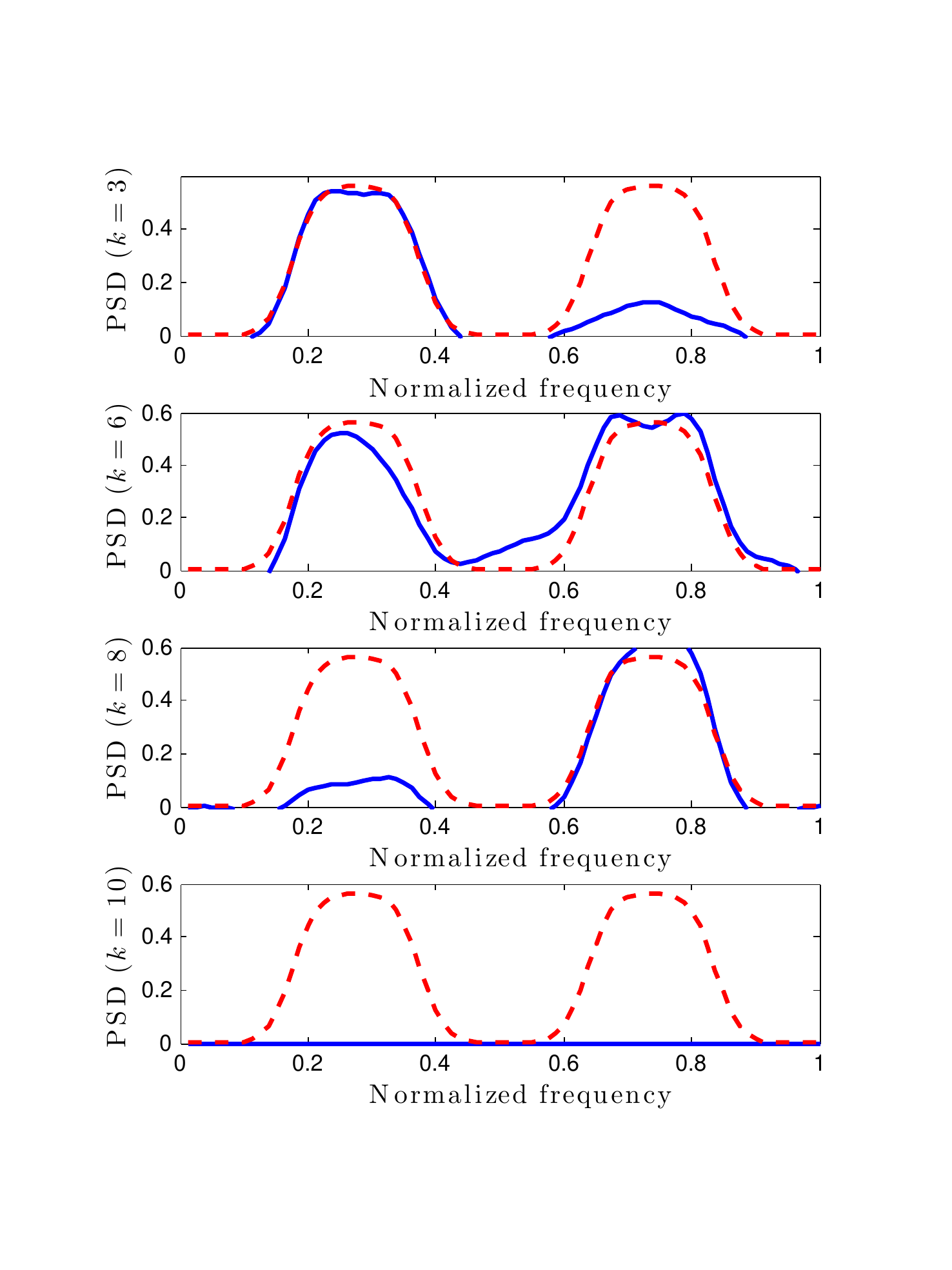}
		\includegraphics[trim = 15mm 25mm 15mm 0mm, clip,scale=0.48]{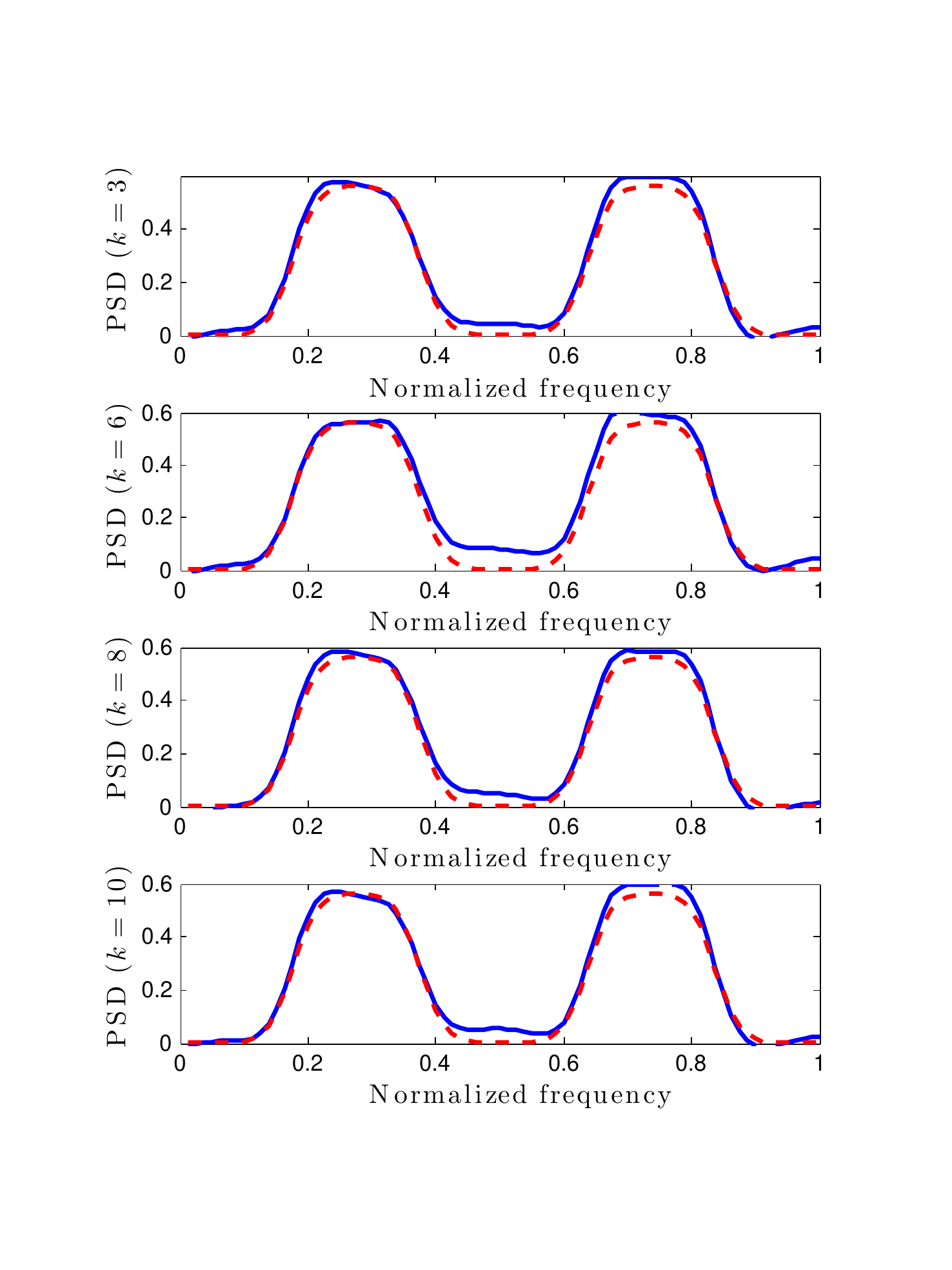}
		\includegraphics[trim = 15mm 25mm 15mm 0mm, clip,scale=0.48]{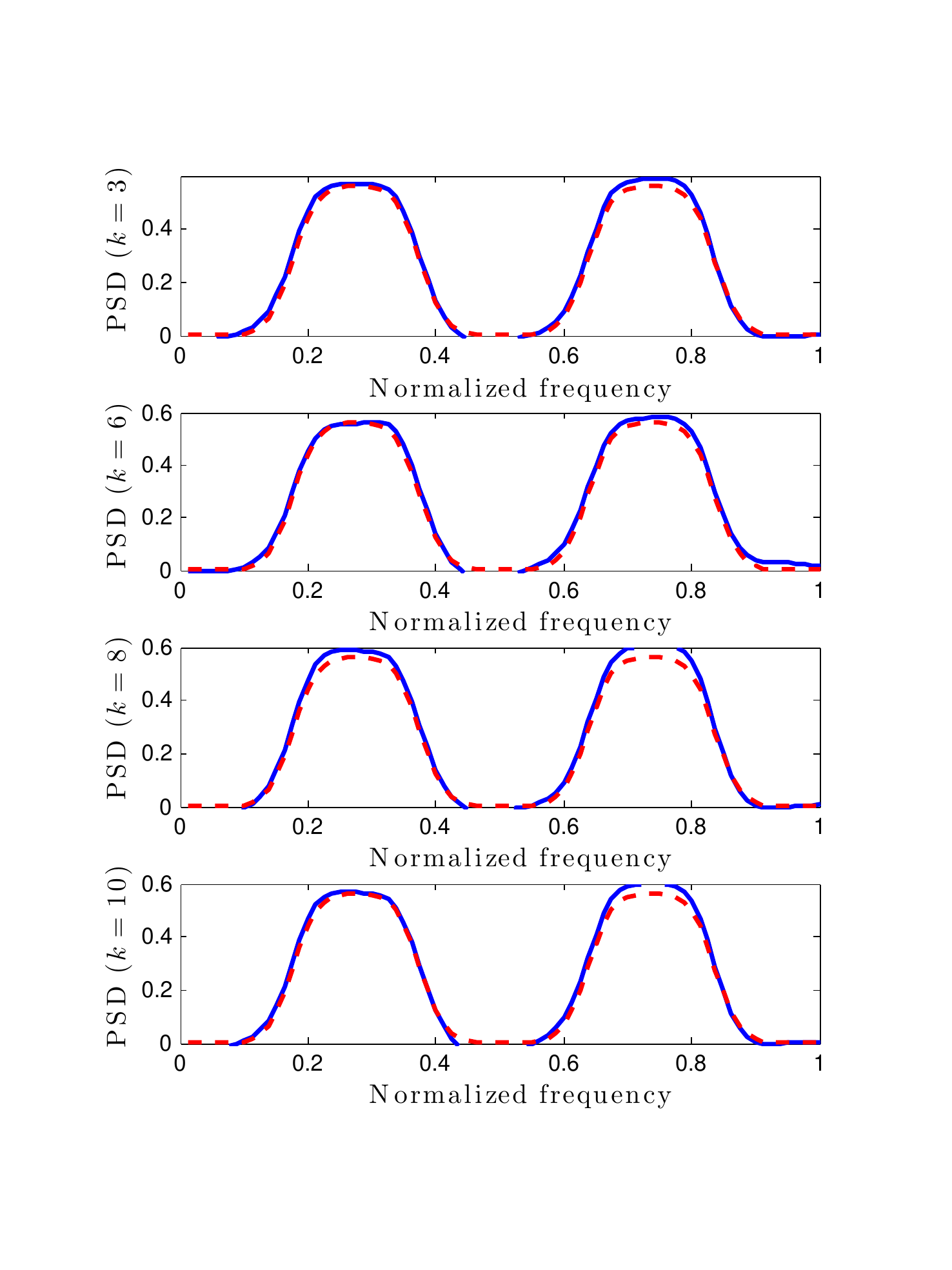}
	\caption{\label{fig:cog_psd}PSD estimation. From top to bottom: estimate on nodes 3, 6, 8, and 10. From left to right: non-cooperative single-antenna
	system ($N_R=1$, $\eta=0$), cooperative single-antenna system ($N_R=1$, $\eta=0.01$), cooperative multi-antenna system ($N_R=4$, $\eta=0.01$).}
\end{figure}

Our algorithm was run on the multi-antenna device network with different settings. In the left plot of Figure~\ref{fig:cog_curve}, a diffusion strategy was applied within each device, without cooperation between devices $(\eta=0)$, for several numbers $N_R$ of antennas per device. In the right plot of Figure~\ref{fig:cog_curve}, a diffusion strategy was applied within each device, and cooperation between devices was promoted $(\eta=0.01)$. The step-sizes were adjusted so that the initial convergence rates were equivalent. It can be observed that increasing the number of antennas, and promoting cooperation between the sets of antennas, allow to improve the performance notably. Figure~\ref{fig:cog_psd} provides the estimated power spectra for $4$ of the $10$ devices. Three learning strategies are considered: a non-cooperative strategy $(\eta=0)$ with single-antenna devices $(N_R=1)$, and a cooperative strategy $(\eta=0.01)$ with single-antenna devices $(N_R=1)$ and multi-antenna devices $(N_R=4)$. 

Red dashed curves represent ground-truth spectra transmitted by the primary users. An inspection of the device locations in Figure \ref{Cognitive_radio} shows that the non-cooperative strategy is highly influenced by the local profile of power spectra. For instance, device $8$ was close to PU1 and far from PU2, and it poorly estimated the power spectra generated by PU2. Device $10$ was not able to estimate any spectrum because it was out of range for PU1 and PU2. Channel allocation relying on such local spectral profile may have led to a hidden node effect. Cooperative strategies clearly provide more consistent results, and using multiple antennas provides additional gain.

\subsection{Distributed non-point target localization}

The second application addresses the problem of target localization. Existing localization methods based on the diffusion strategy assume point targets~\cite{Sayed2013intr,Tu2011Mobile}. However, in some situations, targets may not be reduced to a single point such as its centroid. For instance, this includes the case where the target is a region of interest scanned by a laser light sheet. The algorithm should be able to jointly estimate a series of coordinates that characterizes the target area.

\begin{figure}[!ht]
	\centering
	\includegraphics[trim = 0mm 0mm 20mm 30mm, clip, scale=0.9]{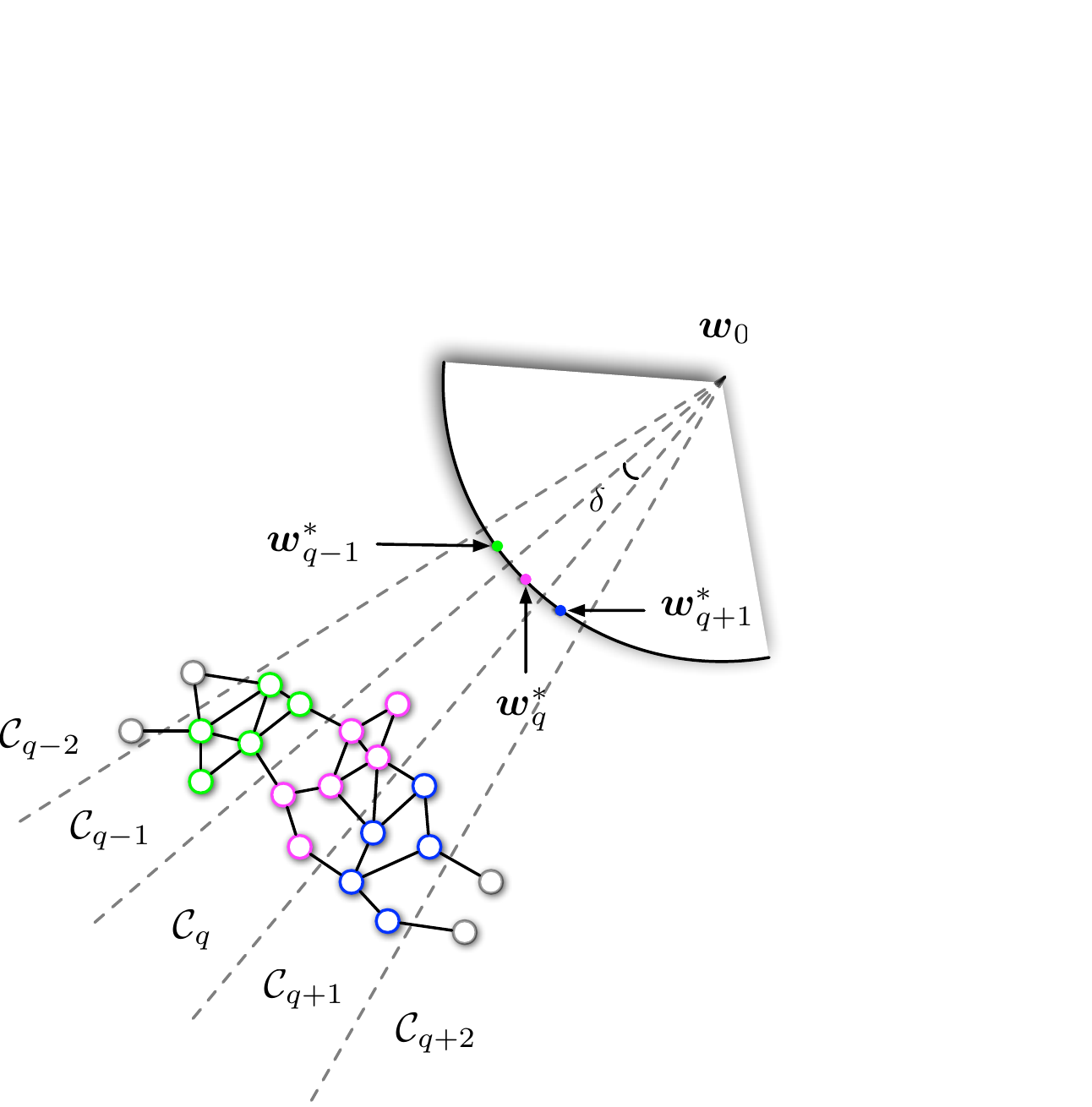}
	\caption{Target surface localization.}
	\label{fig:Source_illustration}
\end{figure}

The problem we considered is shown in Figure~\ref{fig:Source_illustration}. The target was the arc of a circle with center $\bw_o$. The angular resolution of the nodes was denoted by $\delta$. This means that arcs of the circle with solid angle $\delta$ were viewed as a single point $\bw_q$ by the cluster $\C_q$ of nodes within the cone of axis $(\bw_o, \bw_q)$. Note that the distance between each node $k \in \C_q$ and $\bw_q$ can be expressed in the inner product form
\begin{equation}
	r_{kq} = \bu^\top_{kq}\,(\bw_q-\bp_k)
\end{equation}
where $\bp_k$ is the location of node $k$, and $\bu_{kq}$ is the unit-norm vector pointing from $\bp_k$ to $\bw_q$. We assumed that sensors were aware of their location $\bp_k$. Let $d_{kq}=r_{kq}+\bu^\top_{kq}\,\bp_k$, that is, $d_{kq} = \bu_{kq}^\top \,\bw_q$. The problem was thus to estimate $\bw_q^\star$ from noisy input-output data $(\bu_{kq}(n),d_{kq}(n))$ collected by nodes $k\in\C_q$. The model that was thus considered is given by~\cite{Sayed2013intr}:
\begin{equation}
	\begin{split}
	d_{kq}(n) 				&= \bu_{kq}^\top(n) \,\bw_q^\star + v_{kq}(n) \\
	&\text{with}\quad  \bu_{kq}(n) = \bu_{kq} + \alpha_k(n)\,\bu_{kq}^{\perp}+\beta_k(n)\,\bu_{kq}
	\end{split}
 \end{equation}
with $v_{kq}(n)$ a zero-mean temporally and spatially i.i.d. Gaussian noise of variance $\sigma^2_v$. Moreover, the measured direction vector $\bu_{kq}(n)$ was assumed to be a noisy realization of the unit-norm vector pointing from $\bp_k$ to $\bw^\star_q$, with $\alpha_k(n)$ and $\beta_k(n)$ two Gaussian random variables of variances $\sigma_\alpha^2$ and $\sigma_\beta^2$, respectively.

\begin{figure}[!h]
	\centering
	\includegraphics[trim = 20mm 0mm 20mm 0mm, clip,scale=0.45]{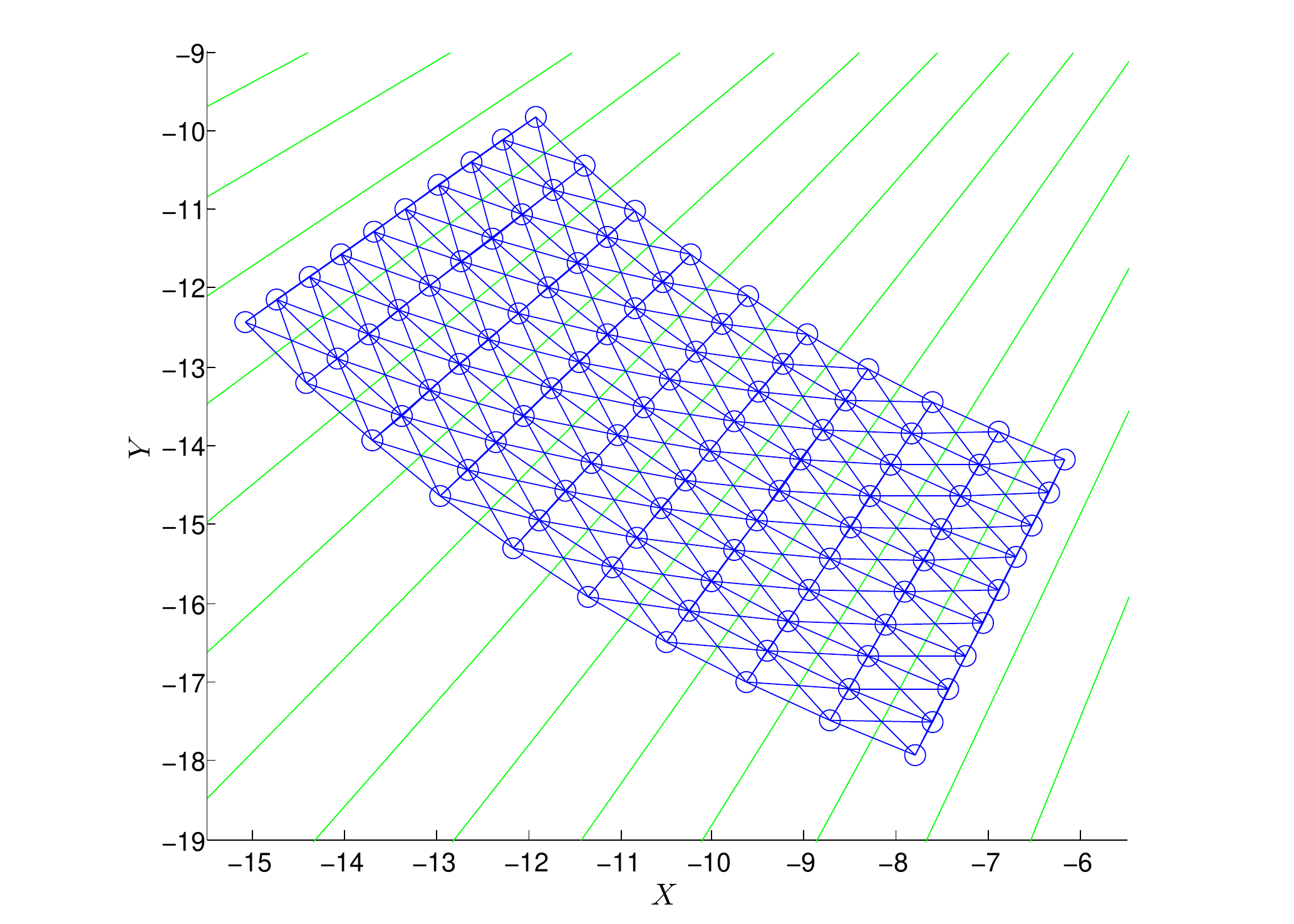}		\qquad
	\includegraphics[trim = 20mm 0mm 20mm 0mm, clip, scale=0.45]{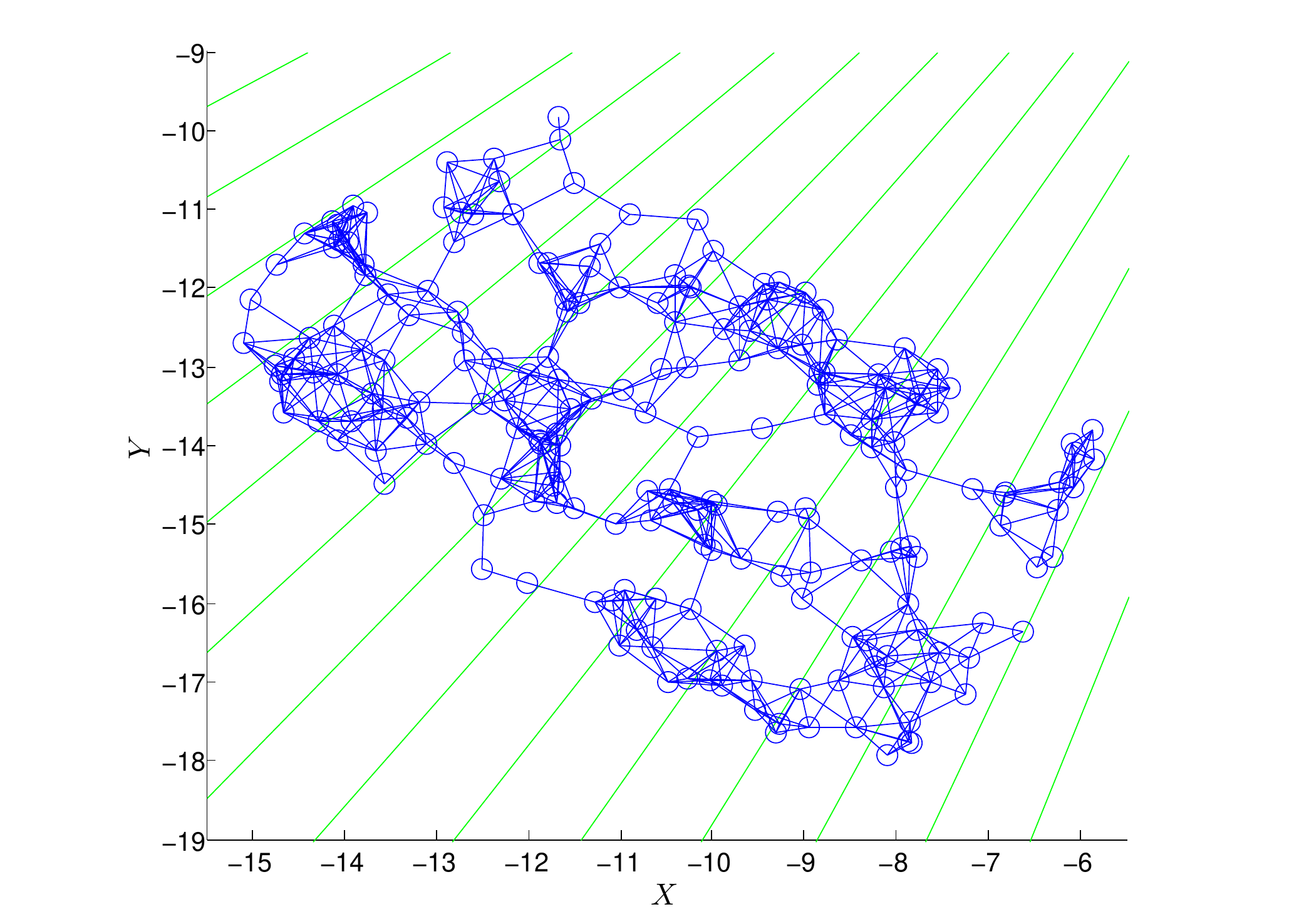}
	\includegraphics[trim = 20mm 0mm 20mm 0mm, clip,scale=0.45]{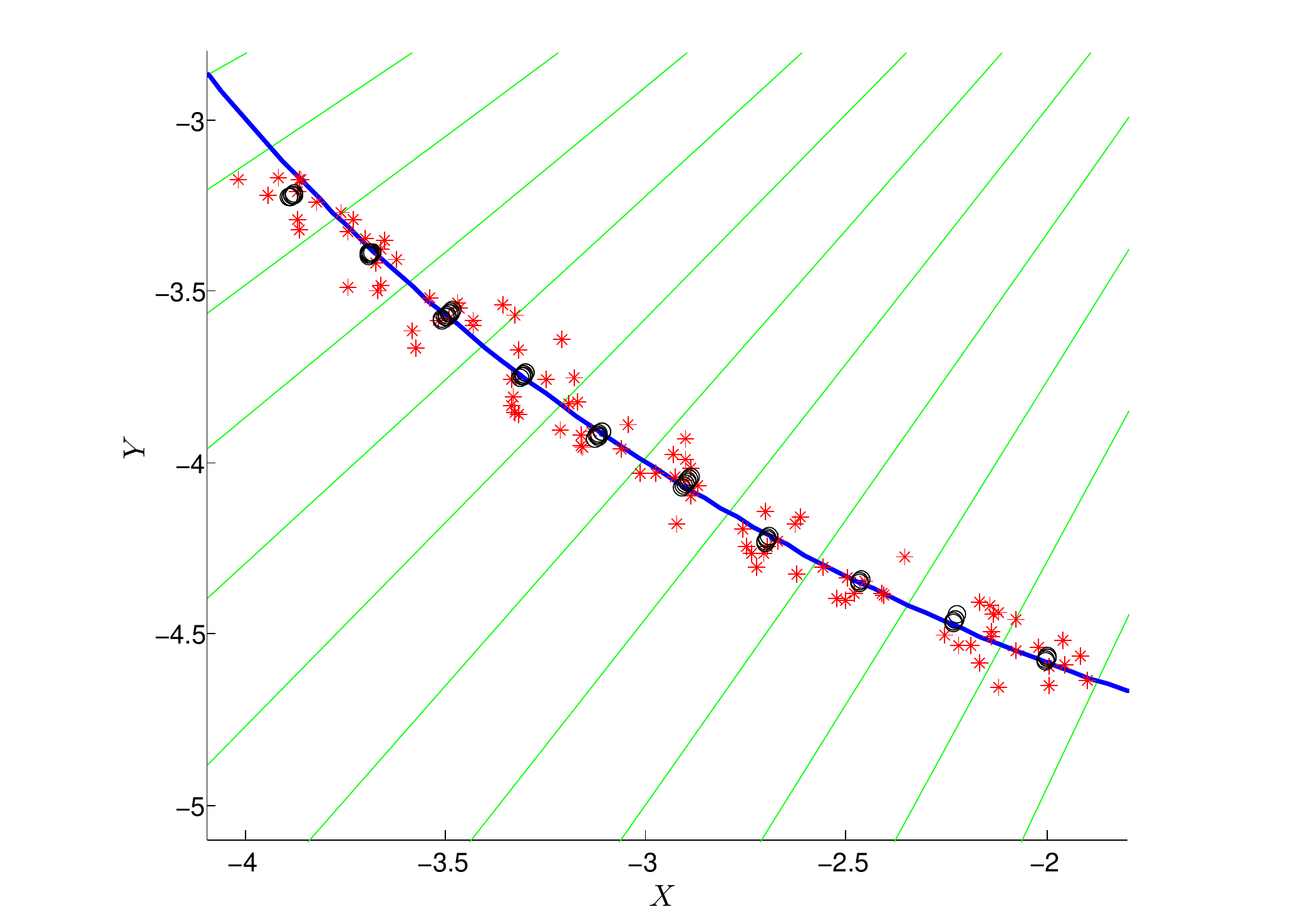}   	\qquad
	\includegraphics[trim = 20mm 0mm 20mm 0mm, clip, scale=0.45]{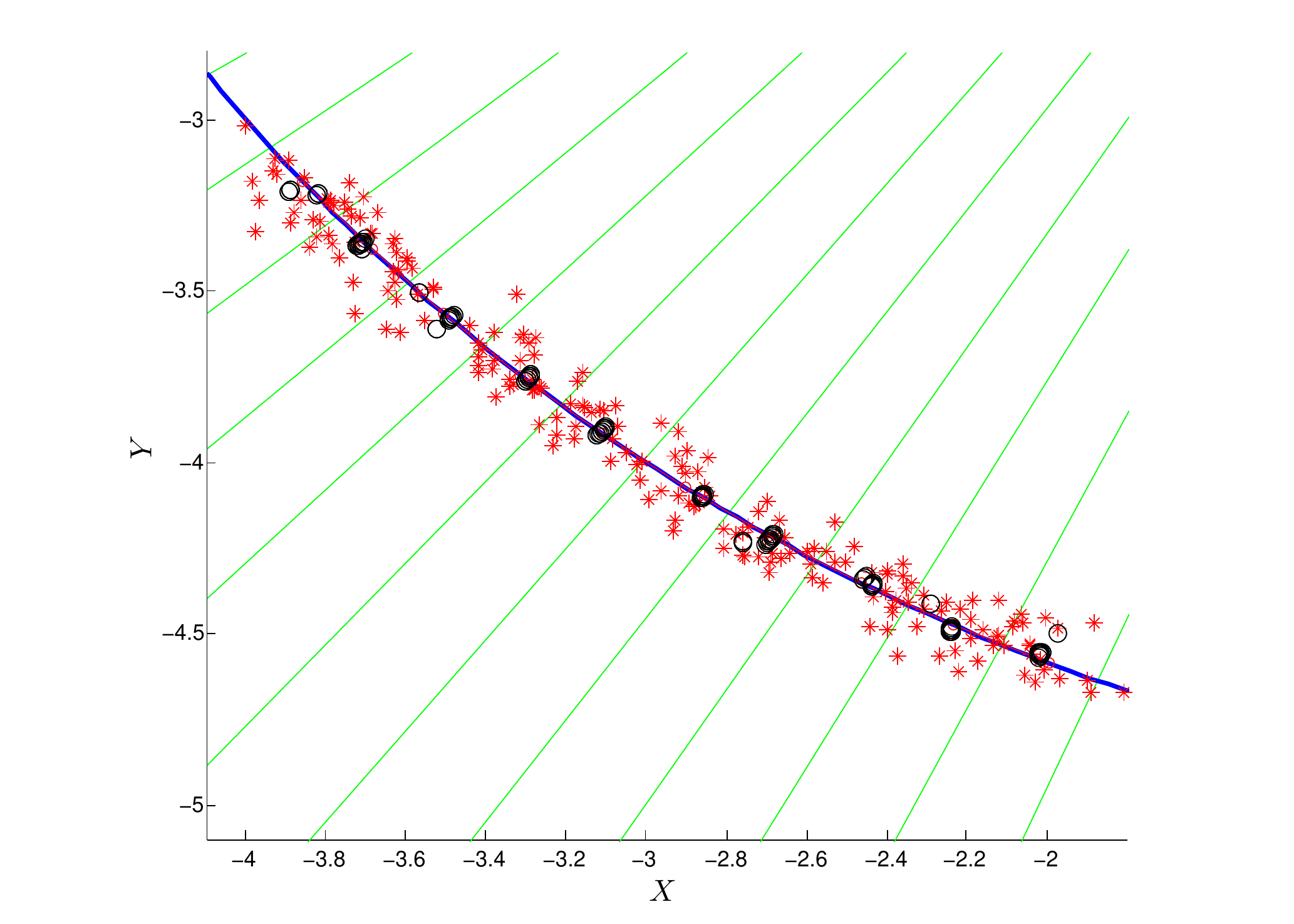}
	\includegraphics[trim = 20mm 0mm 20mm 0mm, clip,scale=0.48]{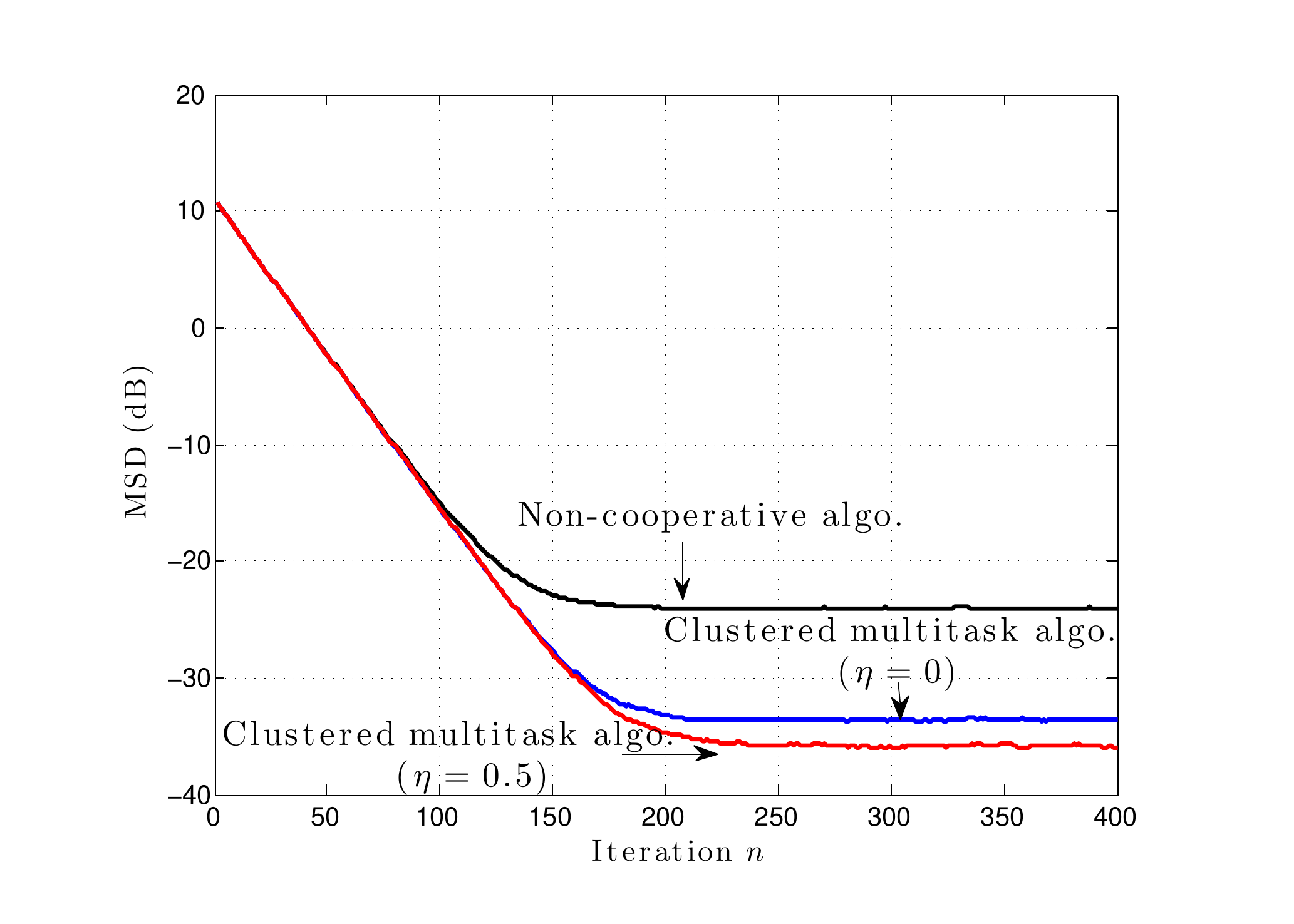}   	\quad
	\includegraphics[trim = 20mm 0mm 20mm 0mm, clip, scale=0.48]{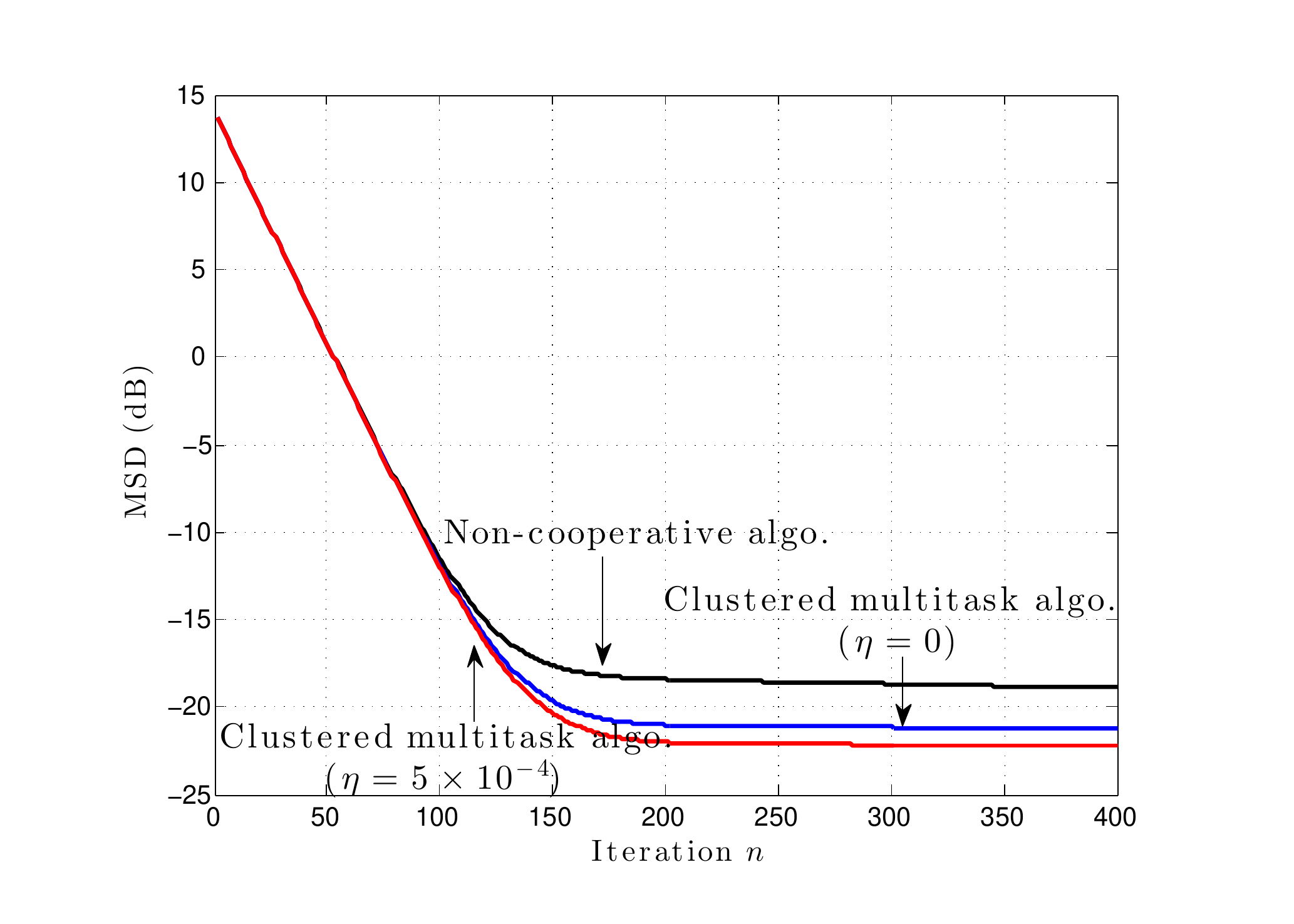}
	\caption{Target surface localization. Left: uniform network. Right: randomly-distributed network.
	Row 1: network connectivity, with cluster boundaries in green. Row 2: estimation results, red crosses for the non-cooperative algorithm,
	black circles for the cooperative algorithm. Row 3: MSD learning curves.}
	\label{fig:source}
\end{figure}

The multitask algorithm \eqref{eq:ATC_MSEl2} was used to estimate the coordinates $\bw_q^\star$ for $q \in \{1, \dots, Q\}$, and to approximate the arc of radius $R$. Each node was connected to its neighbors within its cluster and the adjacent clusters. We considered two network topologies. In the first scenario, see the left-hand plot in Figure~\ref{fig:source} (first row), $100$ nodes ranging from $3R$ to $4R$ were grouped into $10$ clusters, with $10$ nodes in each. The nodes were deployed uniformly with connections between neighbors. In the second scenario, see the right-hand plot in Figure~\ref{fig:source} (first row), $200$ nodes ranging from $3R$ to $4R$ were grouped into $10$ clusters, with $20$ nodes in each cluster. The nodes were deployed randomly. For both experiments, the noise variances were set as follows: $\sigma^2_v=0.5$, $\sigma_\alpha^2=0.1$, and $\sigma_\beta^2=0.01$. We used an identity information exchange matrix $\bC = \bI$. The combination matrix $\bA$ was defined as $a_{\ell k} = |\N{k}\cap\C(k)|^{-1}$ in order to average the estimates of within-cluster neighbors. The regularization strengths $\rho_{k\ell}$ were set to $\rho_{k\ell} = |\N{k}\backslash\C(k)|^{-1}$ for $\ell\in\N{k}\backslash\C(k)$, with $k\neq 1$ and $k\neq Q$. Recall that $\C_1$ and $\C_Q$ are boundary clusters, and the specific regularization strengths $\rho_{1\ell}=\rho_{Q\ell}=0$ for all $\ell$ were used to preserve the configuration of the group.

We ran the non-cooperative algorithm, and the clustered multitask algorithm with $\eta=0.5$ and $\eta=0.0005$ for each scenario, respectively. Figure~\ref{fig:source} (second row) shows one realization of the estimated points $\bw_q$ for each arc. The cooperative algorithm clearly outperformed the non-cooperative algorithm. Figure~\ref{fig:source} (third row) compares the MSD of the two strategies mentioned above, with the clustered multitask algorithm with $\eta=0$. In this case, the diffusion strategy is applied independently in each cluster, without inter-cluster interactions. This experiment clearly illustrates the advantage of fully cooperative strategies in this problem.


 \subsection{Distributed unmixing of hyperspectral data}

Finally, we  consider the problem of distributed unmixing of hyperspectral images using the multitask learning algorithm. Hyperspectral imaging provides 2-dimensional spatial images over many contiguous bands. The high spectral resolution allows to identify and quantify distinct materials from remotely observed data. In hyperspectral images, a pixel is usually a spectral mixture of several spectral signatures of pure materials, termed endmembers, due to limited spatial resolution of devices and diversity of materials~\cite{Keshava2002}. Although nonlinear mixture models have begun to support novel applications~\cite{Bioucas2012,Chen2013-TSP,Chen2013-Whisp}, the linear mixture model is still widely used for determining and quantifying materials in sensed images due to its simpler physical interpretation. With the linear mixture model, pixels can be decomposed as linear combinations of constituent spectra, weighted by fractions of abundance.

To facilitate the presentation, we shall consider that the 3-dimensional hyperspectral image under study has been reshaped into an $L\times N$ matrix $\bY=[\by_1, \dots, \by_N]$, with $N$ the number of pixels and $L$ the number of wavelengths. Let $\bM$ be the $L\times R$ matrix of endmember spectra, with $R$ the number of endmembers, and $\bW=[\bw_1, \dots, \bw_N]$ the $R\times N$ matrix of the abundance vectors of the pixels in $\bY$. The linear mixture model is expressed by
\begin{equation}
	\label{eq:unmixing.problem}
	\bY = \bM \bW + \bV
\end{equation}
where $\bV=[\bv_1, \dots, \bv_n]$  is the modeling error matrix. Suppose that the material signatures (matrix $\bM$) in a scene have been determined by some endmember extraction algorithm~\cite{Winter1999,Nascimento2005,Honeine2011}. The unmixing problem boils down to estimating the abundance vector  associated with each pixel. Besides minimizing the modeling error, it is important to promote similarities of abundance vectors between neighboring pixels due to their possible correlations. Now we write the unmixing problem as follows:
\begin{equation}
	\label{eq:LUM}
	\begin{split}
	&\min_{\bW} \, \|\bY - \bM\bW\|_F^2 + \eta\,\sum_{k=1}^N \sum_{j\in \N{k}} \rho_{kj} \|\bw_k-\bw_j\|_1 \\
	& \text{subject to }\; \bw_{k} \succcurlyeq 0 \quad \text{and} \quad   \cb{1}^\top\bw_k = 1 \quad \text{with} \quad 1 \leq k \leq N,
	\end{split}
\end{equation}
where $\|\cdot\|_F^2$ is the matrix Frobenius norm, $\N{k}$ is the set of neighbors of pixel $k$, $\eta$ is the spatial regularization parameter and $\rho_{kj}$ is the regularization weights. In the above expression, the nonnegativity constraints and sum-to-one constraints are imposed to ensure physical interpretability of the vectors of fractional abundances.

\begin{figure}[t]
	\centering
	\includegraphics[trim = 0mm 0mm 0mm 0mm, clip,scale=0.55]{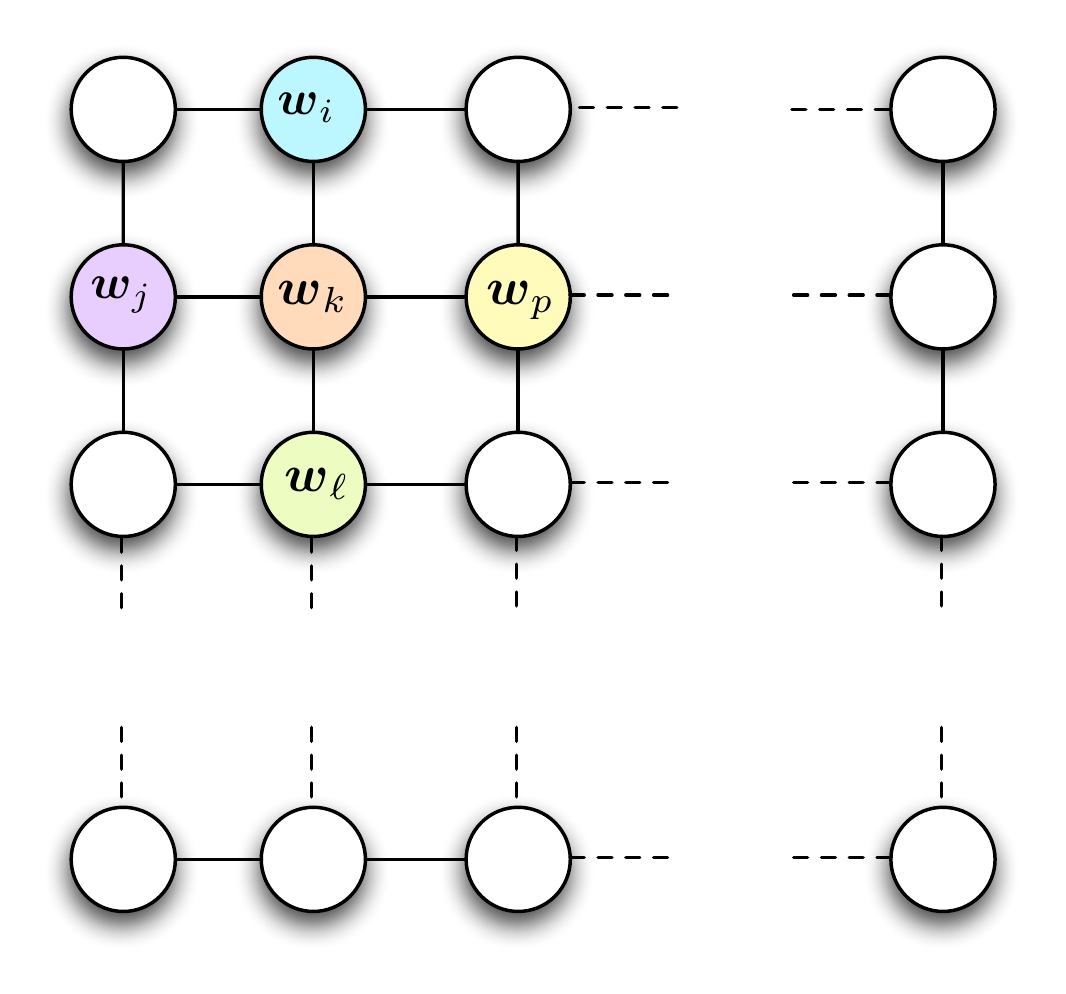}   \qquad
	\caption{  \label{fig:SN_image}Hyperspectral image unmixing problem with first-order connections between neighboring nodes.}
\end{figure}

To conduct linear unmixing of large images in a distributed way, we considered each sensor of the camera as a node, and we applied the diffusion LMS for multitask problems, that is, one node per cluster -- see Figure~\ref{fig:SN_image}. In order to exploit the spatial correlations, we defined the regularization function $\Delta(\bw_k,\bw_j)$ as the $\ell_1$-norm of $\bw_k-\bw_j$ to promote piecewise constant transitions in the fractional abundance of each endmember among neighboring pixels. Similar regularization can be found in~\cite{CHEN2013-TGRS,Chen2013-Icassp}.
This led us to the following algorithm:
\begin{equation}
	\label{eq:LUM_dist}
	\bw_k(n+1) = \cp{P}_{\ell_1^+}\Big(\bw_k(n) + \mu\,\bM^\top(\by_k - \bM\bw_k(n))
	- \mu\,\eta\sum_{j\in\N{k}}\rho_{kj}\,\text{sgn}(\bw_k(n)-\bw_j(n))\Big)
\end{equation}
where we used that the subgradient $\partial_{\bx}\|\bx\|_1=\text{sgn}(\bx)$, with $\text{sgn}(\cdot)$ the component-wise sign function. In this expression, $\cp{P}_{\ell_1^+}(\cdot)$ denotes the iterative operator defined in~\cite{Duchi2008} that projects a vector onto the nonnegative phase of the $\ell_1$-ball to satisfy the nonnegativity and sum-to-one constraint in \eqref{eq:LUM}. This algorithm clearly contrasts with existing batch approaches based on FISTA~\cite{Beck2009fast} and ADMM~\cite{Iordache2012}, which cannot easily address large problems \eqref{eq:unmixing.problem}.

\begin{figure*}[!h]
	\centering
	\includegraphics[trim = 0mm 10mm 0mm 0mm, clip, scale=0.65]{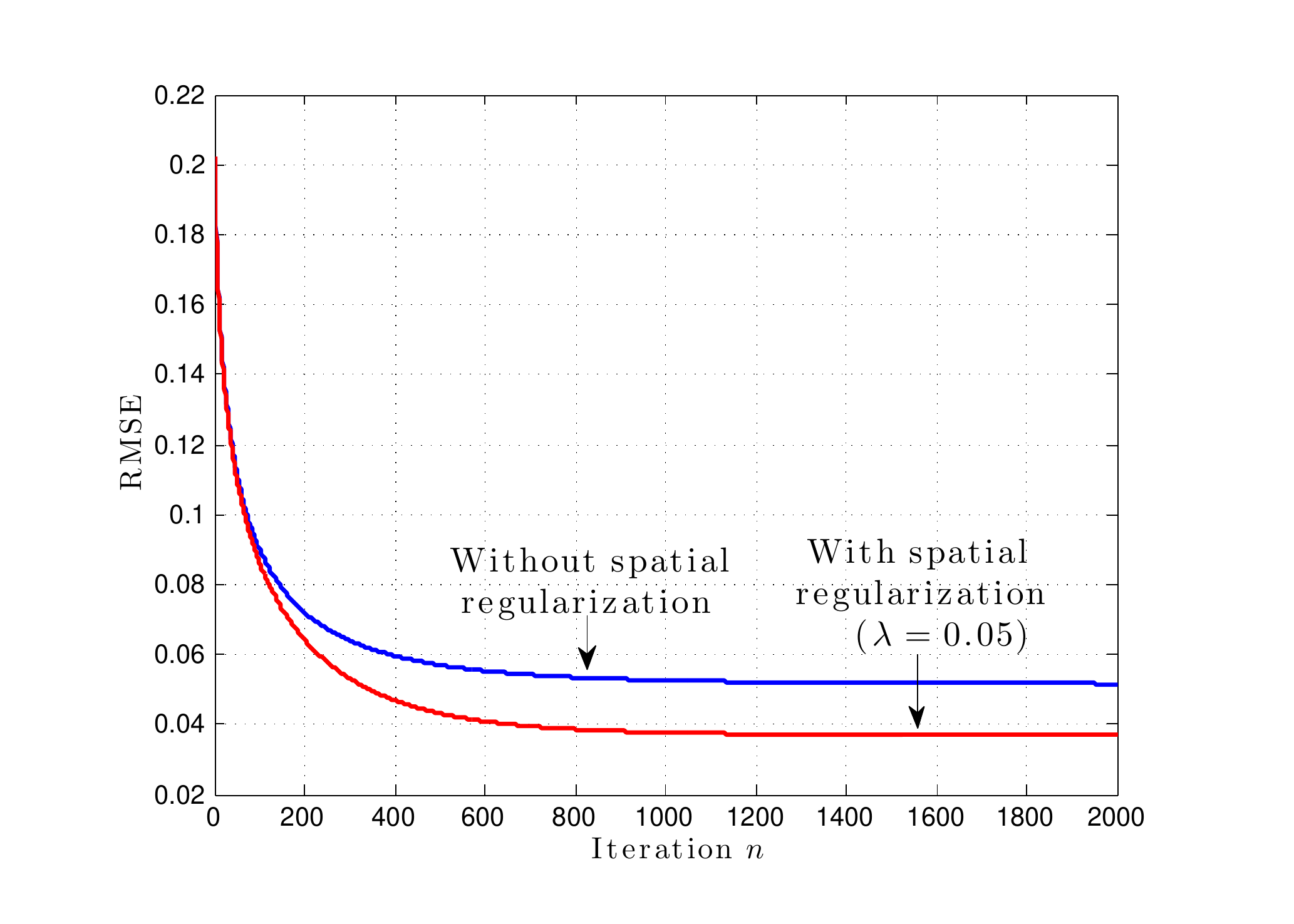}
	\caption{RMSE curve comparison.}
	\label{fig:Hyp_curve}
\end{figure*}

\begin{figure*}[!h]
   		\centering
      		\includegraphics[trim = 0mm 0mm 0mm 0mm, clip, scale=0.8]{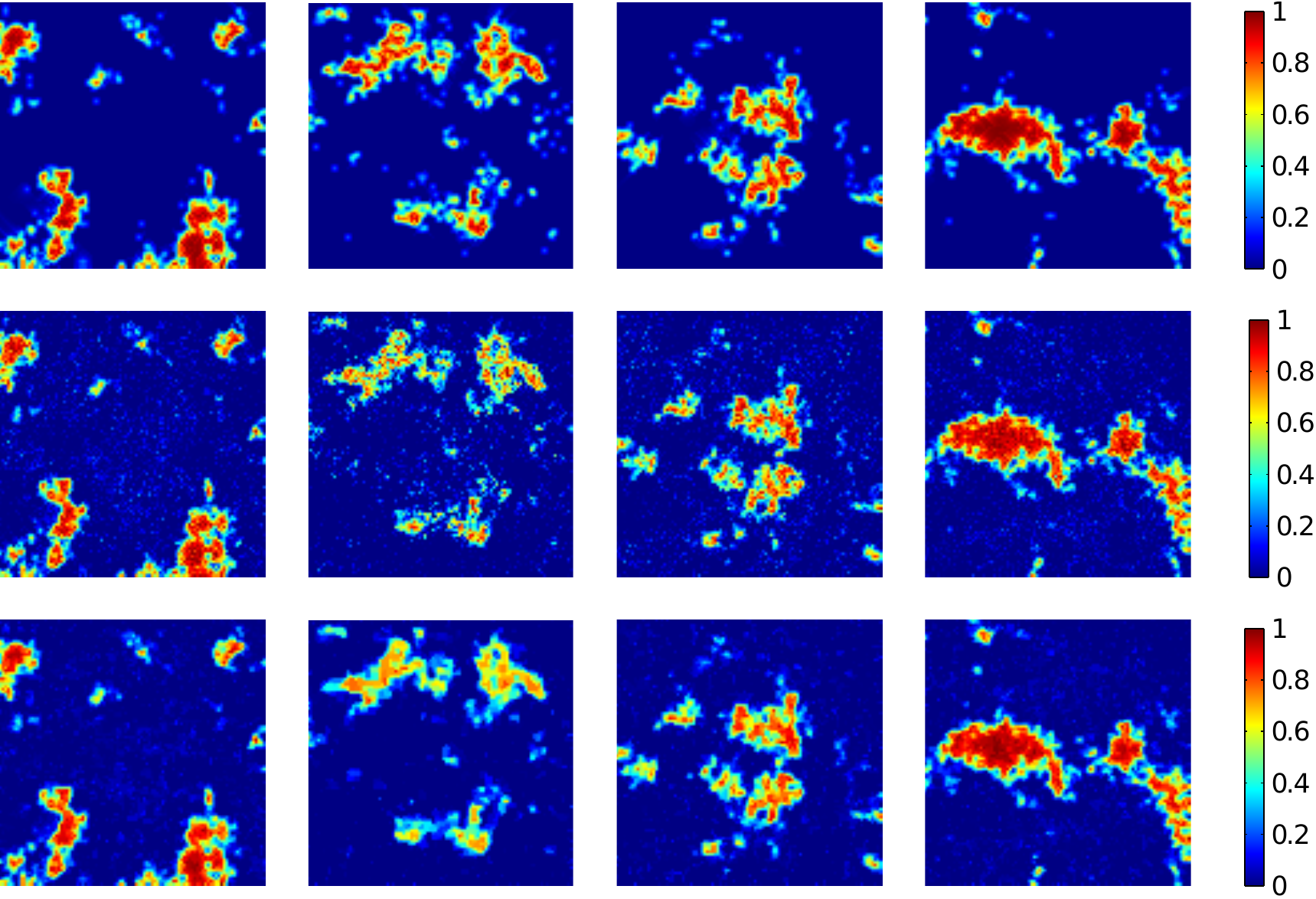}
		\caption{Abundance maps. From left to right: 1st, 6th, 8th, and 9th abundances. From top to bottom: true abundances, estimated abundances without and with spatial regularization.}
		\label{fig:Hyp_imag}
\end{figure*}

The algorithm~\eqref{eq:LUM_dist} was run on a data cube containing $100 \times 100$ mixed pixels. Each pixel was generated by the linear mixture model~\eqref{eq:LUM} using $9$ endmember signatures randomly selected from the spectral library ASTER~\cite{ASTER}. Each signature of this library has reflectance values measured over $224$ spectral bands, uniformly distributed in the interval $3 - 12$ $\mu$m.  The abundance maps of the endmembers are the same as for the image DC2 in~\cite{Iordache2012}. Among these $9$ materials, only the $1$st, $6$th, $8$th, and $9$th abundances are considered for pictorial illustration in Figure~\ref{fig:Hyp_imag}. The first row of this figure depicts the true distribution of these $5$ materials. Spatially homogeneous areas with sharp transitions can be clearly observed. The generated scene was corrupted by a zero-mean white Gaussian noise $\bv_n$ with an SNR level of $20$ dB.  In this experiment, the regularization weights $\rho_{kj}$ were set equal to the normalized spectral similarity:
\begin{equation}
	\rho_{kj} = \frac{\theta(\by_k, \by_j)}{\sum_{\ell\in\N{k}^-} \theta(\by_k, \by_\ell) }
\end{equation}
where $\theta(\by_k, \by_j) = \frac{\by_k^\top \by_j}{\|\by_k\|\|\by_j\|}$. These weights emphasize the regularization between similar pixels and de-emphasize it for less similar pixels. When one knows the ground truth map, a commonly used performance measure for evaluating the performance of an unmixing algorithm is the root mean-square error (RMSE), defined as
\begin{equation*}
	{\text{RMSE}} = \sqrt{\frac{1}{NR}\,\sum_{n=1}^N\|\bw_{n}-{\bw^\star_n}\|^2}.
\end{equation*}
The RMSE learning curves using algorithm~\eqref{eq:LUM_dist}, with spatial regularization ($\eta=0.05$) and without spatial regularization ($\eta=0$), are depicted in Figure~\ref{fig:Hyp_curve}. The corresponding abundance distributions are shown in Figure~\ref{fig:Hyp_imag}. The spatial regularization results in a lower estimation error, and  more homogenous abundance distribution maps with less noise.

%\begin{figure*}[!h]
%	\centering
%	\includegraphics[trim = 0mm 10mm 0mm 0mm, clip, scale=0.65]{Hyp_curve2.pdf}
%	\caption{RMSE curve comparison.}
%	\label{fig:Hyp_curve2}
%\end{figure*}

\section{Conclusion and perspectives}

In this paper, we formulated multi-task problems where networks are able to handle situations beyond the case where the nodes estimate a unique parameter vector over the network. Considering  each parameter vector estimation as a task, and possibly connecting these tasks in order that they can share information, we extended the distributed learning problem from single-task learning to clustered multitask learning. An algorithm was derived. A mean behavior analysis of the proposed algorithm was provided, in the case of the least-mean-square error criterion with $\ell_2$-norm regularization. Several applications that may benefit from this framework were investigated. Several open problems still have to be solved for specific applications. For instance, it would be interesting to show which regularization can be advantageously used with our distributed multitask algorithm, and how they can be efficiently implemented in an adaptive manner. It would also be interesting to investigate how nodes can autonomously adjust regularization parameters to optimize the  learning performance and how they can learn the structure of the clusters in real-time.

\bibliographystyle{IEEEbib}
\bibliography{ref}

\end{document}